\documentclass[10pt,twocolumn,letterpaper]{article}

\usepackage[pagenumbers]{cvpr} % To force page numbers, e.g. for an arXiv version

\usepackage{graphicx}
\usepackage{amsmath}
\usepackage{amssymb}
\usepackage{booktabs}
\usepackage{algorithm}
\usepackage{algorithmicx}
\usepackage{algpseudocode}
\usepackage{multirow}
\usepackage[table,xcdraw]{xcolor}
\usepackage[pagebackref,breaklinks,colorlinks]{hyperref}

\usepackage{booktabs}

\usepackage[capitalize]{cleveref}
\crefname{section}{Sec.}{Secs.}
\Crefname{section}{Section}{Sections}
\Crefname{table}{Table}{Tables}
\crefname{table}{Tab.}{Tabs.}

\def\cvprPaperID{****} % *** Enter the CVPR Paper ID here
\def\confName{CVPR}
\def\confYear{2022}

\begin{document}
\def \sn {\textsc{APTSHIELD}}
\title{\sn{}: A Stable, Efficient and Real-time APT Detection System for Linux Hosts}

\def\eg{e.g\onedot}
\def\etc{etc\onedot}
\definecolor{redcol}{rgb}{1, 0, 0}
\newcommand{\red}[1]{\textcolor{redcol}{#1}} 

% https://tex.stackexchange.com/questions/9594/adding-more-than-one-author-with-different-affiliation
\newcommand*{\affaddr}[1]{#1} % No op here. Customize it for different styles.
\newcommand*{\affmark}[1][*]{\textsuperscript{#1}}
\newcommand*{\email}[1]{\texttt{#1}}
\author{Tiantian~Zhu, Jinkai Yu, Tieming Chen*, Jiayu Wang, Jie Ying, Ye Tian, Mingqi Lv, Yan Chen, \\Yuan Fan and Ting Wang}

\maketitle

%%%%%%%%% MAIN TEXT

\begin{abstract}
    Advanced Persistent Threat (APT) attack usually refers to the form of long-term, covert and sustained attack on specific targets, with an adversary using advanced attack techniques to destroy the key facilities of an organization. APT attacks have caused serious security threats and massive financial loss worldwide. Academics and industry thereby have proposed a series of solutions to detect APT attacks, such as dynamic/static code analysis, traffic detection, sandbox technology, endpoint detection and response (EDR), etc. However, existing defenses are failed to accurately and effectively defend against the current APT attacks that exhibit strong persistent, stealthy, diverse and dynamic characteristics due to the weak data source integrity, large data processing overhead and poor real-time performance in the process of real-world scenarios.
    
    To overcome these difficulties, in this paper we propose \sn{}, a stable, efficient and real-time APT detection system for Linux hosts. In the aspect of data collection, audit is selected to stably collect kernel data of the operating system so as to carry out a complete portrait of the attack based on comprehensive analysis and comparison of existing logging tools; In the aspect of data processing, redundant semantics skipping and non-viable node pruning are adopted to reduce the amount of data, so as to reduce the overhead of the detection system; In the aspect of attack detection, an APT attack detection framework based on ATT\&CK model is designed to carry out real-time attack response and alarm through the transfer and aggregation of labels. Experimental results on both laboratory and Darpa Engagement show that our system can effectively detect web vulnerability attacks, file-less attacks and remote access trojan attacks, and has a low false positive rate, which adds far more value than the existing frontier work.
\end{abstract}

\section{Introduction}
\label{sec:intro}
Advanced Persistent Threat Attacks are escalating to the harm of the current society, and it is often organized by groups of hackers with certain national, governmental or other organizational backgrounds---the hackers are usually well-organized, well-targeted, highly skilled and aggressive, often against the government, core infrastructure (e.g., energy, transportation, communication) and key industries (e.g., military, finance, health care). APT attacks can pose a huge security threat, including confidential data leakage and system integrity damage. APT attacks have occurred frequently in recent years, showing a high incidence of high risk in the world, such as: Stuxnet worm attack in Iranian nuclear power plant \cite{Stuxnet}, BlackEnergy virus attack in Ukraine's power system \cite{BlackEnergy}, and the user information leakage attack in Target \cite{Target}. The COVID-19-themed attacks have also been frequently reported \cite{APTreport}, e.g., APT groups have tried to attack firms working on COVID-19 vaccines and they have used spear-phishing emails to entice users to download and execute malicious attachments so as to steal target-related data and destroy medical infrastructure \cite{covid}.

APT detection has become an important research topic widely concerned with academia and industry field. A series of traditional attack detection schemes such as static code analysis \cite{bolton2017apt,laurenza2017malware}, dynamic sandbox detection \cite{liu2019research,rosenberg2017deepapt}, malicious traffic analysis \cite{zhao2015detecting,huang2020detection}, and hooking technology \cite{mirza2014anticipating,kharaz2016unveil} can be applied to fight against APT attacks. However, the detection effect of the above methods is not ideal due to the problems such as weak de-obfuscate ability, high computational overhead, and low system stability in the practical use. Furthermore, with the extension of the network boundaries and the increment of various 0-day vulnerabilities, it is almost impossible to continue to use these traditional methods to detect APT attacks. Therefore, the idea of Detection and Response (DR) came into being, which is divided into Endpoint Detection and Response (EDR) and Network Detection and Response (NDR). In EDR, the relevant security agent will collect and analyze activity data of the application program when it is running in the user's host/endpoint \cite{EDR}. While NDR will monitor how threats enter the network and how they move laterally in the network \cite{NDR}. Both EDR and NDR can automatically respond to identified threats to remove or contain them. 

Nowadays, the market for EDR solutions is expanding rapidly to meet the urgent need for more efficient endpoint protection and potential vulnerability detection. To this end, the DARPA Transparent Computing (TC) program has tried to organize multiple scientific research units to conduct engagements to enable the prompt detection of APTs and other cyber threats \cite{Engagement5}. EDR tools usually record a large number of system events to a central database. By adopting techniques such as indicators of compromise (IOC), behavior analysis, and machine learning to analyze data, it is hoped that threats will be detected and responded to at an early stage. 

However, in the life cycle of APT attacks, the current EDR system (especially the EDR system of the Linux host) does not form a comprehensive and effective solution to the problems of selecting data sources for detection, massive data analysis and storage, association of suspicious behaviors in context, detection of 0-day threats, as well as energy consumption and real-time performance in detecting APT attacks. The main challenges of the current EDR system based on the Linux host are as follows:

\noindent \textbf{(1) How to select reliable, stable and semantically rich data sources.}
 To carry out complete and accurate APT detection and forensics analysis in the enterprise environment, it is necessary to retain the complete log data of the entire enterprise for subsequent analysis. There are three main log collection methods in the existing attack detection work: The first is the \textit{network data flow}, analysts only need to use a switch that supports port forwarding and a server that analyzes network traffic to obtain relevant data. However, encryption technologies are widely used by sophisticated attackers, making the effective information less available in network data; The second is the \textit{application log}, in this case, both instrumentation and collecting logs directly from the application are limited to the application itself. Analysts cannot observe a global view of the whole system. In other words, the analyst cannot associate the information of application level with that of system level. The third is the \textit{endpoint system log}. Nowadays, there are a few studies focusing on the collection and analysis of system events. However, the existing system-based log collection tools are varied, and few work are able to analyze and evaluate these tools under the current Linux system by combing with collecting performance, data integrity, availability and other comprehensive factors.

\noindent \textbf{(2) How to reduce the amount of data required for real-time detection to improve the detection efficiency.} The duration of APT attacks is usually much longer than other attacks. According to the report from Trustwave \cite{trustwave}, the average latency time of APT attacks is about 83 days, and some of them are as long as several years. Massive data of APT attacks have made new requirements for analysts: It is critical to improve real-time detection efficiency and reduce storage overhead by effective data compaction strategy. Some existing data compaction work \cite{lee2013loggc,lee2013high,ma2016protracer,ma2017mpi,kwon2018mci,kwon2016ldx} tries to use fine-grained taint tracking technology to delete redundant events. However, these methods always rely on known software models which are lack of generalization. Although some studies \cite{hossain2018dependence,hassan2018towards,hassan2020tactical} have proposed general data compaction methods based on audit log, their algorithms read the dependency graph composed of long-term log data into memory at one time, resulting in huge computing and memory overhead. By adopting the above methods \cite{hossain2018dependence,hassan2018towards,hassan2020tactical} in the real-world scenario, data compaction and attack detection cannot be guaranteed in a real-time manner.

\noindent \textbf{(3) How to construct a real-time APT detection framework with high adaptability, high coverage, high precision, low false positives and constant memory overhead.} On the one hand, traditional single point intrusion detection methods (e.g., network traffic analysis, software static feature detection, dynamic sandbox detection and hooking technology) cannot have the ability of anti-escape, reliable detection and real-time alarm at the same time under the threat scenarios with the characteristics of persistence, concealment and diversity of APT attacks. On the other hand, the contextual data analysis studies \cite{hossain2017sleuth,milajerdi2019holmes} usually preserve context information via the provenance graph. However, the size of the provenance graph will explode over time due to the long duration of APT attacks, rendering these approaches inevitably suffer from efficiency and memory problems \cite{xiong2020conan}. Furthermore, various new attack channels and carriers emerge one after another in APT, such as web vulnerability attacks, file-less attacks (also known as in memory attacks or living-off-the-land attacks) and remote access trojan attacks, which are difficult to detect in the initial stage of intrusion. The ideal detection scheme should have good expansibility, be able to cover different types of sophisticated APT attacks, effectively define APT related suspicious behaviors, and comprehensively utilize the full contextual attack chain to realize real-time and accurate detection of attacks.

To cope with the above challenges, we propose a stable, efficient and real-time APT detection system for Linux hosts, called \sn{}\footnote{\sn{} for Stable, Efficient and real-tIme APT Detection system for Linux Hosts.}. Firstly, in order to select reliable, stable and semantically rich data sources, we make a comprehensive and detailed evaluation on Linux system log collection tools. Secondly, for the purpose of reducing the storage overhead and improve the detection efficiency, we reduce the amount of log data by means of redundant semantic skipping and non-viable node pruning. Finally, we construct an APT attack detection framework based on ATT\&CK model \cite{attck}, and the framework can carry out real-time attack response and alarm through the transfer and aggregation of labels. In general, the contributions of this paper are as follows:

\begin{itemize}

\item{Different from previous studies, we make a detailed comparison and analysis of the advantages and disadvantages of the existing data source. We deploy multiple indicators and then obtain the optimal data sources for stable APT detection on Linux through performance analysis and comprehensive judgment.}

\item{We use redundant semantics skipping and non-viable entity pruning to improve the efficiency of real-time APT detection and reduce data storage overhead for forensic analysis. Our data compaction methods can be carried out in real-time streaming data to ensure the performance of the detection system. Also, the compaction effect is better than that of existing studies without affecting the accuracy of final results.}

\item{Based on the ATT\&CK model, we constructed an information aggregation framework through system data flow and control flow. With the help of Tactics, Techniques and Procedures (TTP), the atomic suspicious characteristics of system entities and their transmission rules are defined in the framework to realize the aggregation of the contextual information of APT attacks. Different from traditional single-point detection methods, our framework can aggregate the information of the attack chain into specific entities,  and realize the whole chain detection and real-time alarm of APT attacks with constant memory overhead.}

\item{We implement a stable, efficient and real-time APT detection prototype system on Linux and conduct experiments on the dataset of Darpa Engagement as well as the dataset that simulates APT attacks in real-world scenarios. The experimental results show that our system can effectively detect web vulnerability attacks, file-less attacks and remote access trojan attacks in a real-time manner, and has a low false positive rate, which adds far more value than the existing frontier studies.}

\end{itemize}

The remainder of this article is organized as follows. Section~\ref{sec:related} surveys the relevant work from three aspects: data collection, data reduction and APT detection. Section~\ref{sec:background} describes the preliminary background for our work. System design and evaluation are described in Section~\ref{sec:methods} and Section~\ref{sec:design_experiments}. We conclude and discuss the limitations and improvements of \sn{} in Section~\ref{sec:conclusion}.
\section{Related Work}\label{sec:related}

In this section, we review notable studies in APT detection and compare with them in data collection, data reduction and detection performance to highlight the novelty of our approaches.

\noindent \textbf{Data collection.}
In order to carry out practical and effective APT detection, it's crucial to select the suitable data source which meets the characteristics of non-tampering, low resource consumption, and stability. Network data flow \cite{lu2016apt,stojanovic2020apt} are widely used by previous studies for intrusion analysis and detection, but its universality was insufficient to record all attack operations. Furthermore, encryption technology makes it difficult for analysts to obtain effective information. The application data \cite{su2015framework} shows the inherent characteristics of the program when it is running, but the analyst cannot associate the information of application level with that of system level. Also, some studies \cite{wang2014network,jung2014sensitive} use instrumentation to collect taint records, which may cause huge memory overhead. Some logging tools are also used by researchers, such as NanoLog \cite{yang2018nanolog}, Log4j2 \cite{node4j}, Spdlog \cite{spdlog}, Glog \cite{glog}, and BoostLog \cite{boostlog}. Nanolog is a nanosecond logging system, which is 1-2 orders of magnitude faster than other systems, and its throughput is capable of reaching 80 million log messages per second. However, Nanolog is implemented in C++ and it can only operate on some static strings. Although the remaining systems (i.e., Log4j2, Spdlog, Glog, and BoostLog) are able to meet some requirements, such as low resource consumption and fidelity, they are highly coupled with the application itself, and cannot be used as a general log infrastructure. Also, the above systems cannot log kernel events to monitor all user actions. In view of the deficiencies of the above data sources, kernel-based data collection tools are used by researchers, such as Auditd \cite{auditd}, Sysdig \cite{sysdig}, Lttng \cite{lttng}, and Auditbeat \cite{auditbeat}. However, there are few studies that conduct a complete and detailed analysis of these popular data collection tools based on the needs of the real-world environment.

Different from the previous work, in this paper we compare the performance of the above four kernel-based data collection tools when the system is no/full load through experiments. The details will be discussed in Section~\ref{subsec:collect}. Sysdig is able to depict complete kernel data and parameters, but its requirements for CPU and memory are very high due to its its excessive content. The overall performance of Auditd and Lttng is better than Sysdig, but the output of Lttng is the binary stream, and it takes a lot of time to recover the semantic related data. Auditd has many data filtering modules, it can reduce resource consumption through custom filtering strategies, thereby meeting all the requirements of the detection system for data sources. 

% Hence, we choose Auditd as the data collection tool. 

\noindent \textbf{Data reduction.}
Due to the huge amount of data required (the average latency time of APT attacks is about 83 days) to detect APT attacks, it is critical to improve real-time detection efficiency and reduce storage overhead by effective data compaction strategy. To reduce the system events without affecting the APT detection results is a hot research direction in recent years. 

LogGC \cite{lee2013loggc} pioneered the idea of garbage collection for audit log. The authors of LogGC combined with BEEP \cite{lee2013high} to delete events that had no lasting impact on the system. Subsequently, NodeMerge \cite{tang2018nodemerge} presented a template-based data compaction system with the assistance of an improved FP-Growth algorithm. Some specific patterns such as image loading and system configuration were extracted for data compaction. Similarly, Conan \cite{xiong2020conan} prefiltered the duplicated read events through semantic recognition to reduce the detection efficiency. Zhu et al \cite{zhu2021general} maintained a long list to record redundant events. However, the above three methods are only applicable to specific event types, and have limitations in complex scenarios (e.g., when files are accessed by multiple processes, or there are a large number of file write operations, or the real-time performance of the system will deteriorate as the collection time increases). 

ProTracer \cite{ma2016protracer} tried to improve the compaction rate by dividing the target program into multiple units to perform fine-grained taint tracking. Later, MPI \cite{ma2017mpi} improved ProTracer by utilizing a semantics aware program annotation and instrumentation technique to partition execution, which was able to generate execution partitions with rich semantic information. To avoid application instrumentation or kernel modification, MCI \cite{kwon2018mci} utilized LDX \cite{kwon2016ldx} to acquire precise causal models for a set of primitive operations. The compaction effect of above methods depends on a large number of software models. However, there are a lot of unpredictable software running in the real-world production environment, and it is very difficult to ensure the coverage of these methods. In addition, software updates may invalidate the above methods.

Xu et al. \cite{xu2016high} presented the concept of trackability. By aggregating events under the same trackability (i.e., deleting multiple equivalent data streams and retaining only one of them), they could reduce a large potion of data while preserving events relevant to a forensic analysis. However, this method only considers the characteristics of a single node instead of the global semantics, which makes the compaction effect very limited. To solve this problem, 
both Hossain et al. \cite{hossain2018dependence} and Hassan et al. \cite{hassan2020tactical} presented the data compaction algorithm based on the global semantics of the provenance graph. But during the calculation, the whole data of the provenance graph needs to be read into memory at once, which brings additional I/O and memory overhead. Furthermore, the above methods cannot process real-time streaming data and guarantee the real-time performance of APT detection.

In contrast, \sn{} is able to compact (i.e., redundant semantics skipping and non-viable node pruning) the real-time streaming data to improve the efficiency of real-time APT detection, reduce data storage overhead for forensic analysis, and achieve a better compaction effect than existing work without affecting the accuracy of final result. 

\noindent \textbf{APT detection.}
APT attackers usually exploit zero-day vulnerabilities to compromise the target and conduct long-term penetration. Traditional attack detection schemes such as static code analysis \cite{bolton2017apt,laurenza2017malware}, dynamic sandbox detection \cite{liu2019research,rosenberg2017deepapt}, malicious traffic analysis \cite{zhao2015detecting,huang2020detection}, and hooking technology \cite{mirza2014anticipating,kharaz2016unveil} are not ideal when facing with the characteristics of persistence, stealth, diversity, and clear goals in APT attacks. In order to effectively monitor APT attacks, endpoint detection and response (EDR) is gradually used by researchers. The provenance graph is an ideal method for threat modelling with the ability of powerful semantic expression and attacks historic correlation in the EDR system \cite{li2021threat}. Provenance graph generally records the coarse-grained data from the kernel level, including subjects (e.g., processes and sockets), objects (e.g., files), and edges (e.g., system events). 

BackTracker \cite{king2003backtracking} and PriorTracker \cite{liu2018towards} tried to use provenance graph to identify the entry point and the effect of the attack through backward tracking and forward tracking, respectively. Sun et al. \cite{sun2018using} utilized Bayesian Networks to identify zero-day attack paths on the provenance graph. Similarity, NoDoze \cite{hassan2019nodoze} performed attack triage within the provenance graph to find anomalous paths. During the provenance-based attack investigation, analysts usually came across the dependence explosion problem (i.e., an output event is assumed to be causally dependent on all preceding input events, and an input event is assumed to have causal influence on all subsequent output events). To mitigate this problem, Ma et al. \cite{ma2016protracer} presented a lightweight provenance tracing system to reduce the memory overhead via unit-based execution partitioning \cite{lee2013high}, as well as their improved studies \cite{ma2017mpi,ma2018kernel}. As discussed in data reduction, the above three methods mainly rely on software models and are not universal. 

Due to the persistence of APT attacks, it is difficult for a security analyst to pick out “needle-in-a-haystack” attacks. \textit{Learn-based APT detection} on long-term provenance graphs has been proposed by some researchers. Barre et al. \cite{barre2019mining} tried to extract statistic features of key processes, by adopting a random forest model, their system was able to deliver a detection rate of about 50\%. The results show that simple feature engineering without considering the context of the provenance graph cannot effectively characterize complex APT attacks. Berrada et al. \cite{berrada2019aggregating} designed a comprehensive experiment on provenance graphs by combining existing techniques for APT anomaly detection and they found that simple score or rank aggregation techniques were effective at improving detection performance. Han et al. \cite{han2020unicorn} presented UNICORN which could model system behaviors via structured provenance graphs with a graph sketching technique. Although UNICORN requires no prior expert knowledge of APT attack patterns or behaviors, the anomaly-based system will undoubtedly introduce a large number of false alarms, which is difficult to be practical in real-world scenarios.

In fact, \textit{rule-based APT detection} is more in line with commercial needs, it has been shown that rule-based EDR systems are suitable for addressing the noise problem \cite{numbergame} (i.e., false alarms and duplicate alerts). Expert policies have been proposed by Sleuth \cite{hossain2017sleuth} and Holmes \cite{milajerdi2019holmes} for attack reconstruction on provenance graphs. To match possible exploits of localized components in the provenance graph, empirical labels and prior specifications were used by Sleuth and Holmes, respectively. However, the size of the provenance graph would explode over time due to the long duration of APT attacks, rendering these two approaches inevitably suffer from efficiency and memory problems. Poirot \cite{milajerdi2019poirot} tried to detect APT attacks by measuring the offline similarity between a provenance graph and a query graph. The query graph was obtained based on the expert knowledge from cyber threat intelligence (CTI). The disadvantage of Poirot is that only the occurred attacks are able to be further detected. It is a critical limitation given that composing the elaborate description of a new category of APT requires significant forensics efforts. Conan \cite{xiong2020conan} detected APT attacks by a three-phase model and pre-defined rules. Although Conan is capable of capturing the suspicious behaviour in a real-time manner, it has the following two problems. First, in the case of a large number of concurrent operations in the system (e.g., a large number of file read and write operations at the same time), the real-time performance of Conan cannot be guaranteed because the event generation rate of the target host may be much larger than the event throughput of the detection system. Second, the three-phase model proposed by Conan only reflects the path of suspicious data/control flow, and it has no performance on the tactics used by attackers. That is to say, it lacks interpretable guidance for the analysis of security analysts.

Unlike the previous methods, \sn{} can detect long-term APT attacks with a constant level of memory overhead. By adopting the data compaction strategies, \sn{} is capable of guaranteeing the real-time performance. Also, \sn{} is able to cover different types of sophisticated APT attacks (i.e., web vulnerability attacks, file-less attacks and remote access trojan attacks), effectively define APT related suspicious behaviors based on ATT\&CK model, and comprehensively utilize the full contextual attack chain to realize accurate detection of attacks.

\section{Background}\label{sec:background}

In this section, we first introduce system entities and system events, followed by the dependency graph composed of them. Then, we describe the dependency relationship in the dependency graph. Finally, we introduce how to use provenance graph for forensic analysis.

\subsection{System Entity and System Event}

Generally, the system entity is able to be divided into two categories: subject and object. The subject is the initiator of the event. In most cases, the subject refers to the process. Specifically, users are treated as subjects when there is a user-related event. The object is the target of the event, such as files, network IPs, and channels. It is worth mentioning that processes (e.g., child processes) can also be treated as objects. System events usually record the actions initiated by the subject to the object, such as file reading, file writing, process creation, etc. 

\subsection{Dependence Graph and Dependency}

In order to facilitate efficient analysis of system logs, system entities and events are usually displayed with a dependence graph \cite{king2003backtracking} visually. The dependence graph is also known as a provenance graph or a dependence graph. In these graphs, the nodes including subject and object, and the edges represent the event relationship between subject and object. Dependency graph is a directed graph, and the direction of the event is from subject to object. For each edge, there will be a timestamp and an event name to represent the event sequence and the meaning of the event, respectively. In some research studies of data compaction \cite{lee2013loggc, xiong2020conan, zhu2021general}, system events are able to be merged with multiple timestamps or time ranges. 
% In addition, although the direction of the edge already reflects the direction of the information flow. But the specific types of events should also be recorded on the edge, such as file read, file write, load image, etc.
% But in most cases, the direction of edge is the same as that of information flow, which represent the flow relationship between different entities, and the direction that one entity influences another entity through information flow. 

Dependency graph $G$ can be represented as a combination of $(V, E)$, where V represents all nodes (system entities) and $E$ stands for all edges (system events). For any edge $e \in E$, there is  $e = (u, v, t)$, where $u,v \in V$, $u$ represents the subject, $V$ represents the object, and $t$ represents the time stamp of the event. For the two edges in the dependency graph, $e1=( u1,v1,t1)$, $e2=(u2,v2,t2)$, we consider that there is a dependence between $e1$ and $e2$ if $v1 = u2$ and $t1 < t2$.

\subsection{Forensic Analysis}

Forensic analysis is also known as trace analysis, and its main purpose is to help analysts understand when, how and what impact the attack has been made. In the process of APT analysis, analysts have two tasks. The first one is to determine the entrance of the attack (e.g., the initial process and file entering the terminal and the source IP/port of the attack). The analysis process to achieve this goal is called backward analysis, which is tracing suspicious information flow from the process marked as suspicious through the opposite direction of information flow. Backward analysis was first proposed by BackTracker \cite{king2003backtracking}. During backward analysis, the timestamp is used to judge the causal relationship between different events. For the second task, forward analysis is used to analyze the impact of attacks \cite{zhu2021general, hossain2018dependence}, such as accessing to sensitive information, tampering with system configuration, etc. The origin of forward analysis is usually the attack entry point obtained by backward analysis or any point in the attack chain.
\section{Methods}
\label{sec:methods}

In this section, we first introduce the threat model and the overall architecture of our proposed APT detection system. We then discuss several important topics in its design, including data collection, data compaction, and APT detection framework.

\subsection{Threat Model}

Based on famous APT attack, RSA SecurID tokens leakage \cite{secureid}, we considers the following attack scenarios: an employee received an email with the words "recruitment plan". The employee downloaded and opened the attachment, then was hit by the latest 0-day vulnerability. At the same time, the attacker established a command\&control connection with the employee's host, continuously downloaded and executed the remote access Trojan that the attacker had prepared, which was not able to be detected via the anti-virus software. As a result, the attacker now had access to the target host in the corporate environment. Further, the attacker used this machine as a springboard to perform lateral movement. And the other employee's or senior manager's host that associated with the employee's were successively controlled by the attacker. Then the attacker could stay for months or even years until he completed the ultimate goal, such as the theft of confidential documents or the destruction of the integrity of the system.

In the threat model of this article, the logs generated by the operating system are considered credible. We assume that the system is not attacked before the start of data collection. Attackers can compromise the target host in a variety of ways, including but not limited to 0-day vulnerabilities, social engineering, virus-carrying USB flash drive, etc. The primary task of the attacker is to run the malicious code on the victim's machine, and then collect target information through remote command\&control. Attackers will hide themselves in normal system activities through disguise, with the ultimate goal of stealing high-value data or destroying the integrity and availability of the system.

\subsection{System Overview}

\begin{figure*}
    \centering
    \includegraphics[width=2\columnwidth]{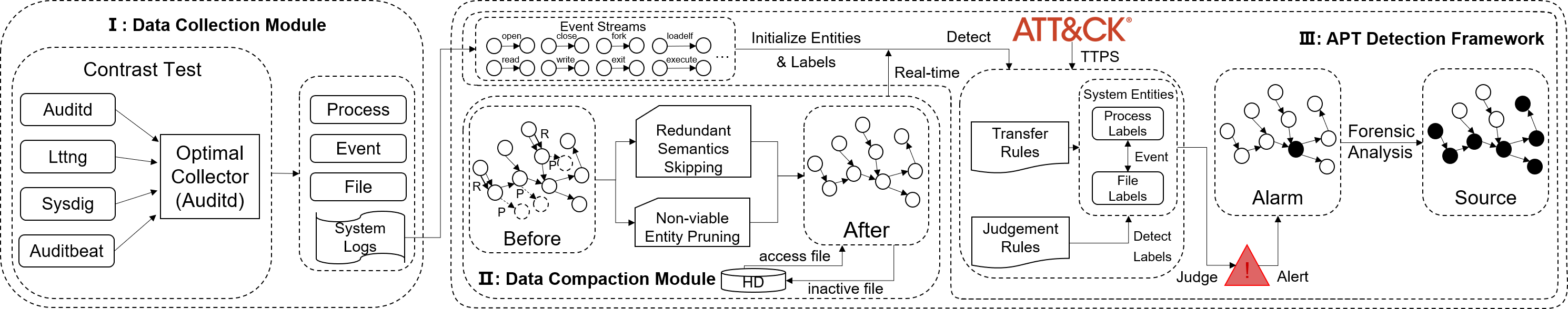}
    \caption{System Overview of \sn{}. Data collection module, data compaction module and APT detection framework are the key components of \sn{}. The kernel-based tool (Auditd) is deployed on the client side for system logs collection. By adopting redundant semantics skipping and nonviable node pruning, data compaction module will help data storage overhead. APT detection framework is responsibility for real-time aggregation of the contextual information of APT attacks with the help of ATT\&CK model. Once the threat/attack occurs, \sn{} can find the entire attack chain through forensic analysis.}
    \label{fig:systemoverview}
\end{figure*}

The architecture of \sn{} is illustrated in Figure \ref{fig:systemoverview}. It consists of three parts: data collection module, data compaction module, and APT detection framework. Firstly, kernel-based tools are adopted for data collection. Through comprehensive analysis, we choose auditd as the collector on the client side since it provides sufficient data with low overhead. The collected system logs (including entities and events) will construct a dependence graph. Secondly, redundant semantics skipping and nonviable node pruning will be adopted to compact system logs. The main advantage of data compaction is to improve the efficiency  of real-time APT detection and reduce data storage overhead for forensic analysis. Thirdly, APT detection framework is proposed based on the ATT\&CK model. The atomic suspicious characteristics of system entities (i.e., process labels and file labels) and their transfer rules (through events with different semantics) are defined in the framework to realize the aggregation of the contextual information of APT attacks. The APT will be alerted from a specific entity (through judgement rules), and finally the related attack chain will be found through forensic analysis.

\subsection{Data Collection}\label{subsec:collect}

In order to detect APT in a real-time manner with high efficiency, it is crucial to select optimal data sources and data collectors. To this end, in this section we first sort out the data sources for APT detection under the Linux system, and then analyze and select the best data collector that meets the requirements. 

\subsubsection{Data Requirements}

According to the ATT\&CK model, the host-based APT detection system needs to collect relevant data from different stages (e.g., initial access, untrusted execution, data exfiltration, etc.). We summarize requirements of both data collector and data sources into the following three aspects:

\noindent \textbf{(1) Tamper-proof.}
APT attacks are camouflaged. In order to avoid detection, APT attackers often hide their attack traces (e.g., bash history and log message can be easily modified/deleted by an attacker, thereby increasing the difficulty of attack detection). We need to select credible data sources to ensure that intruders cannot avoid detection by tampering with data. 

\noindent \textbf{(2) Stability and Low Overhead.}
APT attacks are persistent. That means data collection is a long-term process. Firstly, the data collector needs to be directly deployed on the user’s host. If the data collection takes up too much resources, it will affect the user’s normal usage. Secondly, the detection system needs to perform stable and real-time detection and alarm, this requires data to be collected and transmitted in real-time without any data loss. Thirdly, the operation of the data collector cannot affect the stability of the system (e.g., it should not be requested to modify the kernel) 

\noindent \textbf{(3) Rich semantics.}
APT attacks are diverse. It is necessary to analyze a series of data such as file IO data, network IO data, inter-process communication data, process attribute modification data, sensitive system operation data, and file attribute modification data. 

Based on three requirements above, the kernel-level logs are used as the main data source. We analyze and select the best data collector currently available on Linux, which will be described in details in the next section. 

\subsubsection{Data Source Analysis and Selection}

As introduced in Section~\ref{sec:related}, current data collectors that are active in the community are sysdig, auditd, lttng, and auditbeat. With the Intel Xeon E5 CPU (8 cores) and 64G memory running on Ubuntu 16.04, we analyze and compare the performance (CPU and memory usage) of these data collectors when the host is no-load and full-load (i.e., use commands to increase the CPU usage to more than 95\%. The executed commands include but are not limited to file reading and writing, file downloading, inter-process communication, executing shell scripts, accessing pages, etc.), respectively. The results are shown in Table~\ref{tab:data_source_comparison}.

% \begin{table}[h]
% \centering
% \caption{Performance comparison of four different kernel-based data collectors. We tested the CPU (single core) and memory usage of each collectors on full-load and no-load. All the results are the average based on 10 tests independently.}
% \label{tab:data_source_comparison}
% \resizebox{\columnwidth}{10mm}{
% \begin{tabular}{|c|c|c|c|c|}
% \hline
% \multicolumn{1}{|l|}{} & \textbf{no-load CPU} & \textbf{no-load RAM} & \textbf{full-load CPU} & \textbf{full-load RAM} \\ \hline
% Auditd    & 0.2\%  & Related to the buffer & 2.0\%  & Related to the buffer \\ \hline
% Lttng     & 2.0\%  & 413252K               & 23.0\% & 478780K               \\ \hline
% Sysdig    & 19.0\% & 111028K               & 22.0\% & 474564K               \\ \hline
% Auditbeat & 1.2\%  & 139688K               & 7.1\%  & 174512K               \\ \hline
% \end{tabular}}
% \end{table}

\begin{table}[]
\centering
\caption{Performance comparison of four different kernel-based data collectors. We tested the CPU (single core) and memory usage of each collectors on full-load and no-load. All the results are the average based on 10 tests independently.}
\label{tab:data_source_comparison}
\begin{tabular}{|c|c|c|c|c|}
\hline
\multirow{2}{*}{\textbf{Collector}} & \multicolumn{2}{c|}{\textbf{No-load}}                                                                                        & \multicolumn{2}{c|}{\textbf{Full-load}}                                                                                      \\ \cline{2-5} 
                                    & \textbf{\begin{tabular}[c]{@{}c@{}}CPU\\ (\%)\end{tabular}} & \textbf{\begin{tabular}[c]{@{}c@{}}Memory\\ (MB)\end{tabular}} & \textbf{\begin{tabular}[c]{@{}c@{}}CPU\\ (\%)\end{tabular}} & \textbf{\begin{tabular}[c]{@{}c@{}}Memory\\ (MB)\end{tabular}} \\ \hline
\textbf{Auditd}                     & 0.2                                                         & 31.2                                                    & 2.0                                                         & 31.2                                                    \\ \hline
\textbf{Lttng}                      & 2.0                                                         & 403.6                                                          & 23.0                                                        & 467.6                                                          \\ \hline
\textbf{Sysdig}                     & 19.0                                                        & 108.4                                                          & 22.0                                                        & 463.4                                                          \\ \hline
\textbf{Auditbeat}                  & 1.2                                                         & 136.4                                                          & 7.1                                                         & 170.4                                                          \\ \hline
\end{tabular}
\end{table}

It can be seen from Table~\ref{tab:data_source_comparison} that auditd and auditbeat have better overall performance when the server is fully loaded. The overhead (memory) of auditd is the same under both no-load and full-load because its memory usage is related to the buffer size. In the actual test, the buffer size can be controlled within 40MB to meet the data demand under full load. If sysdig is used as a collector, there will be service crashes and data loss when the system is fully loaded. The output data generated by lttng is a binary byte stream, and parsing requires a lot of costs. Considering that auditd is a module that comes with the kernel and has the lowest cost, we finally select auditd as the data collector in this article. In addition, we use the stress testing tool UnixBench \cite{unixbench} to monitor the state of auditd when it is turned on/off. The results show that when auditd is turned on, the total score of UnixBench drops about 0.3\% compared to that is turned off, which is basically negligible. 

% In order to further verify that auditd can satisfy the requirements of our detection work. We also used unixbench to perform a stress test on the state of auditd when it is turned on and off. The following Table \ref{tab:unixbench_test} shows that, after auditd is turned on, the system call consumption score of unixbench has dropped from 2138381 to 2088508, and the bandwidth of the pipe file has dropped from 2230349 to 2161036. These two performances dropped by about 3\% after opening auditd, and the total score dropped from 1643 to 1638, which is basically negligible.

% \begin{table}[h]
% \centering
% \caption{Stress test on the state of auditd by Unixbench. The higher the score of each item, the better the performance. The score of the last item reflects the overall situation.}
% \label{tab:unixbench_test}
% \resizebox{\columnwidth}{6mm}{
% \begin{tabular}{|c|c|c|c|c|}
% \hline
% \textbf{State} & \textbf{Bandwidth of Pipe File} & \textbf{Process Creation} & \textbf{System Call} & \textbf{Score} \\ \hline
% Off            & 2230349.5                       & 4400.6                    & 2138381.2            & 1643.3         \\ \hline
% On             & 2161036.9                       & 4477.1                    & 2088508.8            & 1638.3         \\ \hline
% \end{tabular}}
% \end{table}

\subsection{Data Compaction}

If the collected raw data is sent directly to the server for processing, it will consume a lot of storage and network resources due to the long duration of the APT attack. Therefore, in the real-world enterprise environment, these data should be compacted first, and then sent to the storage/detection server. The data compaction method should meet the following conditions: \noindent \textbf{(1) Real-time.} The data consumption rate should be greater than the data generation rate to avoid the impact of cached data on the real-time performance of the detection system; \noindent \textbf{(2) High efficiency.} The data compaction method has low CPU and memory usage, and will not affect the operating system itself. Also, it will greatly reduce the data transmission bandwidth and data storage cost; \noindent \textbf{(3) High accuracy.} The compacted data should maintain the dependencies in the original data, and should not have a negative impact on the APT detection and forensic analysis of APT attacks. In this section, we will first give the insight for data compaction. Then we introduce two algorithms in our data compaction method: redundant semantics skipping and non-viable node pruning.

\subsubsection{Insight for Data Compaction}

In order to perform data compaction without affecting the detection results, it is important to analyze the contextual semantics of dependency graph. In the attack detection based on the dependency graph, it is necessary to analyze the interaction behavior of the system entities by judging the different semantics of the system events (that is, the edges of the dependency graph), and then find the path related to the attack. From the perspective of semantic transfer, we assume that entity $u$ has an event $e1$ directed to entity $v$ at time $t1$. It is considered that entity $u$ hash an impact on entity $v$ at time $t1$. Then at time $t2$ $(t2 > t1)$, there is an event $e2$ pointing to entity $w$ from entity $v$, so it is considered that entity $v$ has an impact on entity $w$ at time $t2$. Since the event $e2$ occurs after the event $e1$, it can be considered that the entity $u$ has an impact on the entity $w$, which is transmitted through the entity $v$.

\begin{figure}
    \centering
    \includegraphics[width=0.7\columnwidth]{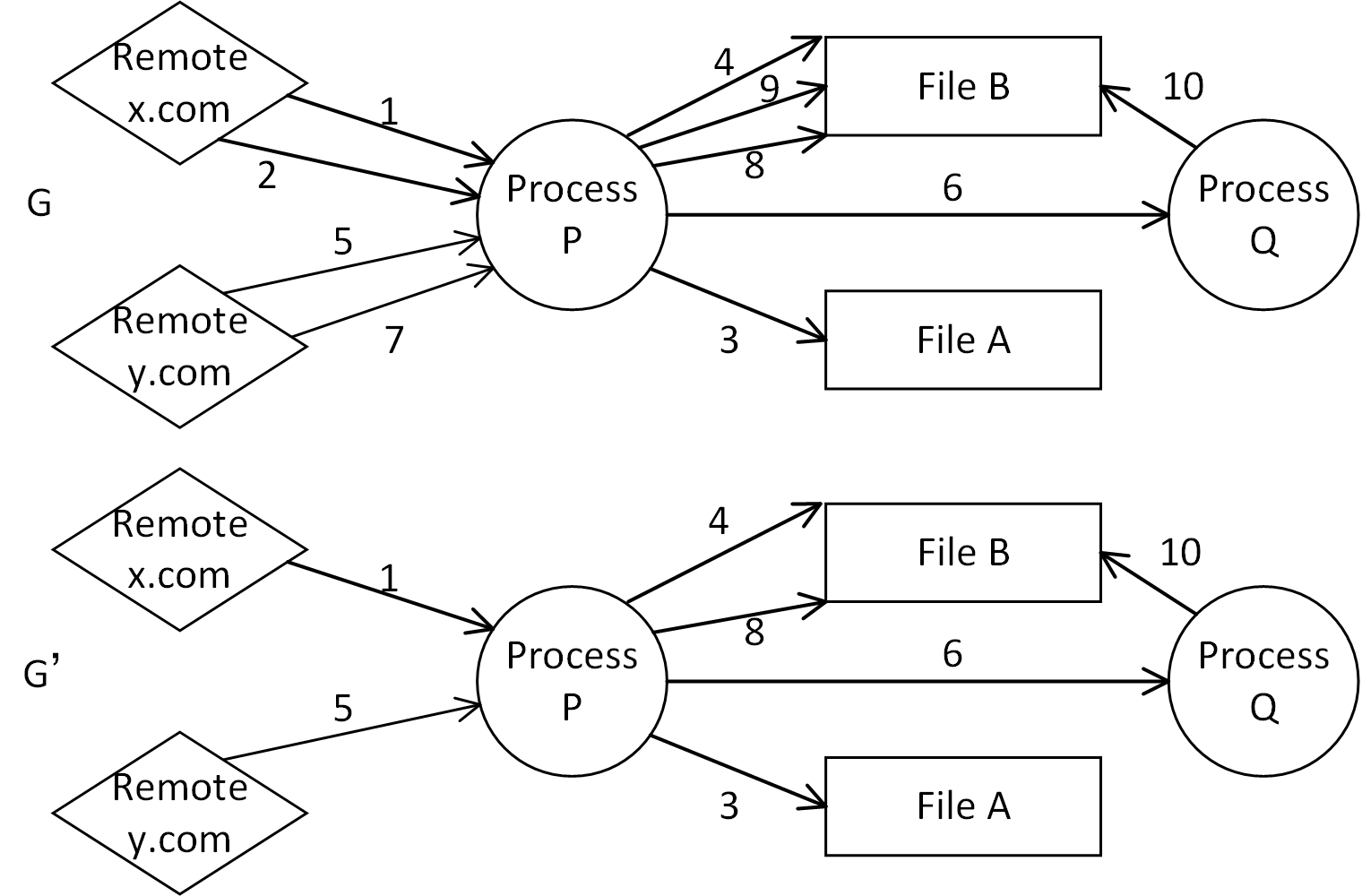}
    \caption{An example about Data compaction. It contains the original provenance graph $G$ and the compacted graph $G'$.}
    \label{fig:data_compaction}
\end{figure}

Take Figure~\ref{fig:data_compaction} as an example. The circle represents the process, the rectangle represents the file, the diamond represents the IP/DNS, the arrow represents the event, and the number on the arrow represents the time point of the events (the smaller the number, the earlier the event occurs). Process $P$ is affected by the event generated by $x.com$ at time $t1$, and $P$ obtains semantic information from the network. After that, the system event with process $P$ as the initial entity will have a direct impact on other entities (i.e., file $A$, file $B$, and process $Q$). However, considering that process $P$ is only affected by the event generated by $x.com$ before time $t5$. It can be judged that from time $t1$ to time $t5$, other entities pointed by process $P$ are only affected by the network information from $x.com$ to $P$. After the event at time $t5$ occurs, other entities pointed by process $P$ are simultaneously affected by the network information from both $x.com$ to $P$ and $y.com$ to $P$. According to observations above, we can see that the impact of the event on the entity is lasting. From the perspective of contextual semantics, when an entity receives an event containing external information, the entity will contain certain characteristics transmitted by the event. And these characteristics will affect the flow of information from this entity. Our main idea/insight of data compaction is derived from these observations: \textbf{If the semantics of source entity has not changed and the destination entity has received multiple same information streams (events) from the source entity, these events will have the same semantics, which can be combined to delete redundant semantics}.

% \begin{figure}
%     \centering
%     \includegraphics[width=0.9\columnwidth]{fig/RedundantSemanticDeletionBefore.jpg}
%     \caption{Before Redundant Semantic Deletion}
%     \label{fig:before_redundant_semantic_deletion}
% \end{figure}

The result of redundant semantics skipping is shown in Figure~\ref{fig:data_compaction}. The data compaction process is as follows:

(1) After the time $t1$, process $P$ already has the semantics from $x.com$, and the event at $t2$ do not introduce any new semantics and can be deleted.

(2) Although during $t5$ and $t7$ an event (at $t6$) from $P$ to $Q$ occurred, it only affects the semantics of $Q$ and the semantics of $P$ remains unchanged. The events at time $t5$ and $t7$ have the same semantics, and we can delete the event at time $t7$.

(3) We can find that events at $t4$,$t8$, and $t9$ have the same subject and object. The event at time $t5$ has changed the semantics of $P$, so the event at time $t8$ should be retained. Since the events at time $t8$ and time $t9$ have the same semantics, we can delete the event at time $t9$.

Note that in this example, we assume that the semantics of IP/DNS (i.e., external network) was not changed. But in the real-world scenario, the external network may be compromised by an attacker, thereby causing semantic changes. To address this problem, we design a time window $T$ to retain the network receiving events at a low frequency, which is inspired by \cite{hossain2018dependence,zhu2021general}. For example, if $T = 50$, then after every 50 time intervals, the semantics of IP/DNS will be refreshed and retained again. We choose $T = 50$ in this paper.

% \begin{figure}
%     \centering
%     \includegraphics[width=0.9\columnwidth]{fig/RedundantSemanticDeletionAfter.jpg}
%     \caption{After Redundant Semantic Deletion}
%     \label{fig:after_redundant_semantic_deletion}
% \end{figure}

\subsubsection{Redundant Semantics Skipping}
\sn{} maintains a table to store the latest semantic information of all processes. When the semantics of the incoming event is the same as that in the semantic table, the event can be considered as redundant and deleted. If there are a large number of repeated read and write operations, deleting redundant semantic events can greatly reduce execution time and memory overhead.

The algorithm of redundant semantics skipping is shown in Algorithm~\ref{alg:gsskip_algorithm}. The input is the real-time streaming data (events) and the output is whether to delete the event or not. In addition, we use $LST$ to record the set of (entity, event) that have been occurred. When processing the real-time event, the algorithm will determine whether the subject $S$ exists in $LST$. If it exists, then determine whether the event $e$ is the same as stored event $e'$. If it is the same, the event will be deleted. We found that when the size of $LST$ gets larger, the efficiency of this algorithm will become lower. Therefore, we propose to empty the list when the $LST$ reaches a certain size. In the test, it is found that when the threshold is set to 5, \sn{} has better performance and efficiency. In addition, when the event $e$ is a WRITE/RECV event, the event $e'$ stored in $LST$ related to $O$ need to be deleted. The reason is that the semantics of the object $O$ will be changed due to the WRITE/RECV event. 

\begin{algorithm}
    \small
    \caption{Data Compaction of Redundant Semantics Skipping}
    \textbf{Input:} \\
    (1)Streaming data (events) in chronological order, each event $e_{i}$ contains a subject $S_{i}$ and a object $O_{i}$\\
    (2)The set of (entity, event) that have been occurred recently are denoted as Latest Semantic Table $LST$\\
    \textbf{Output:} \\
    Whether the event is skipped \\
	\textbf{Initialize:} \\
	The set of (entity, event) $LST = \varnothing$
        \begin{algorithmic}[1]
            \For{$e_{i}$ $\in$ streaming events}:
                \If{$S_{i}$ not exists in $LST.Keys$}
                    \State{$LST$ add ($S_{i}$,$e_{i}$)}
                \Else
                \If{$e_{i}$ equals to $e'_{i}$}
                    \State{Skip the event}
                    \State{Continue}
                \Else
                    \State{Delete ($S_{i}$,$e'_{i}$) in $LST$}
                    \State{Add ($S_{i}$,$e_{i}$) to $LST$}
                \If{$O_{i}$ exists in $LST.Keys$}
                    \State{Delete ($O_{i}$,$e'_{i}$) in $LST$}
                \EndIf
                \EndIf
                \EndIf
                \If{$LST.size$ $\ge$ 5}
                    \State{$LST = \varnothing$}
                    \State{Continue}
                \EndIf
                \If{$e_{i}$ $\in$ $\{$WRITE, RECV$\}$}
                    \For{($S'_{i}$,$e'_{i}$) $\in$ $LST$}:
                        \If{$O_{i}$ exists in $e'_{i}$}
                            \State{Delete ($S'_{i}$,$e'_{i}$) in $LST$}
                        \EndIf
                    \EndFor
                \EndIf
            \EndFor
        \end{algorithmic}
    \label{alg:gsskip_algorithm}
\end{algorithm}

\subsubsection{Non-viable Entity Pruning}
After filtering out redundant semantics, we maintain a node tree to contain the relationship of all processes for efficient forensic analysis. For all process entities that do not have a parent node (i.e., independent processes), we connect them by creating a virtual root process. In order to keep a stable memory overhead when the system runs for a long time, non-viable entity pruning is used to reduce the number of processes. As shown in Algorithm ~\ref{alg:non-survival_entity_pruning_algorithm}, if a process satisfies: (1) it executes the exit event, (2) it does not have potential harmful functionalities (PHF, it will be introduced in Section~\ref{subsec:label_definition}) and child nodes (When the child node has PHFs, the deletion of the parent node will affect the forensic analysis), this process will be pruned from the node tree. 

Moreover, we move inactive file nodes (in this article, inactivity is defined as no change for more than 5 minutes, and the time interval can be changed according to actual needs) to the disk to further reduce the memory overhead. They will be took out from the disk in real-time when needed. 

% The algorithm of non-viable entity pruning is shown in Algorithm~\ref{alg:non-survival_entity_pruning_algorithm}. The input is the real-time streaming data (only EXIT events) and the output is whether the process is pruned. And we use $PT$ to record the process tree. We use the function $isleaf(P)$ to determine whether the process $P$ has child processes and the function $PHF(P)$ to determine whether the process $P$ contains malicious tags, which defined later in 4.5.2. If the process $P$ does not have child processes nor contains malicious tags, the process $P$ will be pruned from the process tree $PT$.

\begin{algorithm}
    \small
    \caption{Data Compaction of Non-viable Entity Pruning}
    \textbf{Input:} \\
    Streaming data (only EXIT events) in chronological order. Each event $e_{i}$ contains a subject $P_{i}$\\
    \textbf{Output:} \\
    Whether the process is pruned \\
	\textbf{Initialize:} \\
	For all processes are stored in a Process Tree $PT$
        \begin{algorithmic}[1]
            \For{$e_{i}$ $\in$ streaming events}:
                \If{$e_{i} \in  \{EXIT\}$}
                    \If{$isleaf(P_{i}) \And \lnot PHF(P_{i})$}
                        \State{prune $P_{i}$}
                    \EndIf    
                \EndIf
            \EndFor
        \end{algorithmic}
    \label{alg:non-survival_entity_pruning_algorithm}
\end{algorithm}

\subsection{APT Detection Framework}

In this section, we first introduce the insight for label-based APT detection. Then we explain the definition of labels and label delivery rules based on ATT\&CK model. Finally we give the attack judgment rules.

\subsubsection{Insight for Label-based APT Detection}

Due to the persistence of APT attacks, the detection methods based on the dependency graph suffer from efficiency and memory problems. To detect long-term APT attacks with a constant level of memory overhead, we develop various adaptable labels based on ATT\&CK model, and transfer contextual semantics between entities through control flow and data flow. \textbf{Control flow} mainly refers to the process creation relationship. When the attack involves multiple processes, tracking the untrusted control flow can realize the association of multiple malicious processes. \textbf{Data flow} refers to the dependence of data content, which reflects the flow path of data. The data flow can be divided into untrusted data flow and high-value data flow. The former one reflects the flow path of external untrusted data, which is often the intrusion path of suspicious codes; the latter one reflects the escape path of high-value contents in the attack after being stolen. Untrusted control flow, untrusted data flow, and high-value data flow reflect the nature of the attack. By tracing the above three information flows, analysts can easily characterize the attack chain. However, when the subject/object of system events are different, events of the same type may have different semantics. For example, we consider the file reading event, reading a downloaded file (R1) represents an untrusted data flow, while reading a text document in a user’s key directory (R2) represents a high-value data flow. The semantics of these two file reading events are not the same. In the ATT\&CK model, R1 is mostly the behavioral characteristic in the Initial Access stage of the APT attack, which may lead to malicious code execution, while R2 may be the Data Exfiltration stage where the attacker is stealing user information, which may eventually lead to user information leakage. In order to solve the problems above, we define a series of entity labels, as well as transfer rules for labels between entities to visually describe the APT attack. In summary, \textbf{\sn{} gathers the semantics of the complete attack chain into target entity to realize the abstraction of large-scale attack features with constant-level memory overhead}.

\subsubsection{Label Definition}\label{subsec:label_definition}

% We use the label delivery method for detection. Label the process and file nodes, because this way we can save the taint information in the label, such as the process reads high-value information or the file contains sensitive information. The following will define the tags used in this article.
\noindent \textbf{(1) Process Labels.}
We divides process labels into two categories: status labels and behavior labels, as shown in Table~\ref{tab:process_label}. The status labels represent the label that semantics in the process will be passed to the child process with the FORK event. The behavior labels indicate what the process has done, and are used to accurately locate the threat. For example, if a process has network connection, we label it with $PS1$, for we can not trust the data from network. If a process executes a command without being allowed, we label this process with $PB4$. A detailed description of the process label is in Appendix~\ref{exp:process_label}.

\begin{table}[h]\Large
\centering
\caption{Process label definition; including status labels and behavior labels.}
\label{tab:process_label}
\resizebox{\columnwidth}{!}{
\begin{tabular}{|c|c|c|}
\hline
\textbf{Labels} & \textbf{Description} & \textbf{Category} \\ \hline
$PS1$ & Process has network connection & \multirow{7}{*}{Status} \\ \cline{1-2}
$PS2$ & Process accesses to data in high-value data flow nodes & ~ \\ \cline{1-2}
$PS3$ & Process contains network data & ~ \\ \cline{1-2}
$PS4$ & Process loads or reads the uploaded file from network & ~ \\ \cline{1-2}
$PS5$ & Process interacts with non-existent files & ~ \\ \cline{1-2} 
$PS6$ & Process reads the /etc/passwd file & ~ \\ \cline{1-2}
$PS7$ & Process reads the .bash\_history file & ~ \\ \hline
$PB1$ & Process executes a file from network & \multirow{8}{*}{Behavior} \\ \cline{1-2} 
$PB2$ & Process executes a sensitive file & ~ \\ \cline{1-2}
$PB3$ & Process executes a sensitive command & ~ \\ \cline{1-2}
$PB4$ & Process has executed commands (for Webshell only) & ~ \\ \cline{1-2}
$PB5$ & Process executes commands which can let others get the shell & ~ \\ \cline{1-2}
$PB6$ & Process modifies the /etc/crontab file & ~ \\ \cline{1-2}
$PB7$ & Process modifies the /etc/sudoers file & ~ \\ \cline{1-2}
$PB8$ & Process reads high-value information & ~ \\ \hline
\end{tabular}}
\end{table}

% \begin{table}[h]\tiny%
% \centering
% \caption{Process label definition; including status labels and behavior labels.}
% \label{tab:process_label}
% \resizebox{\columnwidth}{32mm}{
% \begin{tabular}{|c|c|c|}
% \hline
% \textbf{Labels} & \textbf{Description} & \textbf{Category} \\ \hline
% $PS1$ & Process has network connection & \multirow{7}{*}{Status} \\ \cline{1-2}
% $PS2$ & Process accesses to data in high-value data flow nodes & ~ \\ \cline{1-2}
% $PS3$ & Process contains network data & ~ \\ \cline{1-2}
% $PS4$ & Process loads or reads the uploaded file from network & ~ \\ \cline{1-2}
% $PS5$ & Process interacts with non-existent files & ~ \\ \cline{1-2} 
% $PS6$ & Process reads the /etc/passwd file & ~ \\ \cline{1-2}
% $PS7$ & Process reads the .bash\_history file & ~ \\ \hline
% $PB1$ & Process executes a file from network & \multirow{8}{*}{Behavior} \\ \cline{1-2} 
% $PB2$ & Process executes a sensitive file & ~ \\ \cline{1-2}
% $PB3$ & Process executes a sensitive command & ~ \\ \cline{1-2}
% $PB4$ & Process has executed commands(for Webshell) & ~ \\ \cline{1-2}
% $PB5$ & Process executes commands which can let others get the shell & ~ \\ \cline{1-2}
% $PB6$ & Process modifies the /etc/crontab file & ~ \\ \cline{1-2}
% $PB7$ & Process modifies the /etc/sudoers file & ~ \\ \cline{1-2}
% $PB8$ & Process reads high-value information & ~ \\ \hline
% \end{tabular}}
% \end{table}

\noindent \textbf{(2) File Labels.}
We divide file labels into two categories: untrusted labels and high-value labels, as shown in Table \ref{tab:file_label}. The untrusted files refer to the file containing untrusted data from the network, and the high-value files refer to the file containing sensitive data. For example, if a file contains data from the network, we label it with $FU2$, for the untrsted data may cause an attack later (code execution). A detailed description of the file label is in Appendix~\ref{exp:file_label}.

Some processes with specific labels may directly or indirectly cause potential damage to system security. We define them as potential harmful functionalities (PHF). In this paper, we define $PS2,3,5$ and $PB1,2,5$ as PHF. Administrators are able to add more labels to the PHF list to implement strict control. 

\begin{table}[h]\Large
\centering
\caption{File label definition; including untrusted labels and high-value labels.}
\label{tab:file_label}
\resizebox{\columnwidth}{!}{
\begin{tabular}{|c|c|c|}
\hline
\textbf{Labels} & \textbf{Description} & \textbf{Category} \\ \hline
$FU1$ & File is uploaded & \multirow{6}{*}{Untrusted} \\ \cline{1-2}
$FU2$ & File contains data from the network &  ~ \\ \cline{1-2}
$FU3$ & File does not exist & ~ \\ \cline{1-2}
$FU4$ & File is written by the Webshell attack & ~ \\ \cline{1-2}
$FU5$ & File is written by the RAT attack & ~ \\ \cline{1-2}
$FU6$ & File is written by the Living-off-the-land attack & ~ \\ \hline
$FH1$ & File that can control scheduled tasks such as /etc/crontab & \multirow{4}{*}{High-value} \\ \cline{1-2}
$FH2$ & File that can control permissions such as /etc/sudoers & ~ \\ \cline{1-2}
$FH3$ & File that holds sensitive information such as /etc/passwd & ~ \\ \cline{1-2}
$FH4$ & File that saves historical commands such as .bash\_history & ~ \\ \cline{1-2}
$FH5$ & File is written by process that have read sensitive information & ~ \\ \hline
\end{tabular}}
\end{table}

% \begin{table}[h]\tiny%
% \centering
% \caption{File label definition; including untrusted labels and high-value labels.}
% \label{tab:file_label}
% \resizebox{\columnwidth}{28mm}{
% \begin{tabular}{|c|c|c|}
% \hline
% \textbf{Labels} & \textbf{Description} & \textbf{Category} \\ \hline
% $FU1$ & File is uploaded & \multirow{6}{*}{Untrusted} \\ \cline{1-2}
% $FU2$ & File contains data from the network &  ~ \\ \hline
% $FU3$ & File does not exist & ~ \\ \cline{1-2}
% $FU4$ & File is written by the Webshell attack & ~ \\ \cline{1-2}
% $FU5$ & File is written by the RAT attack & ~ \\ \cline{1-2}
% $FU6$ & File is written by the livingoffland attack & ~ \\ \hline
% $FH1$ & File that can control scheduled tasks such as /etc/crontab & \multirow{4}{*}{High-value} \\ \cline{1-2}
% $FH2$ & File that can control permissions such as /etc/sudoers & ~ \\ \cline{1-2}
% $FH3$ & File that holds sensitive information such as /etc/passwd & ~ \\ \cline{1-2}
% $FH4$ & File that saves historical commands such as .bash\_history & ~ \\ \cline{1-2}
% $FH5$ & File is written by process that have read sensitive information & ~ \\ \hline
% \end{tabular}}
% \end{table}

\subsubsection{Event Selection}
After defining the labels of system entities (i.e., processes and files), it is necessary to clarify effective events between entities in order to facilitate the transfer and aggregation of contextual information. Inspired by \cite{hossain2017sleuth,xiong2020conan}, the events involved are a part of the events that exist in the Linux kernel data collected by auditd, including process events such as fork, execute, LoadELF, file operation, entity attribute modification, network operation, etc.. A list of the events we used is in Appendix~\ref{exp:event_used}.

% \begin{table}[h]
% \centering
% \caption{Event Definition}
% \label{tab:event_label}
% \begin{tabular}{|c|c|}
% \hline
% \textbf{Events} & \textbf{Description}\\ \hline
% E0 & File Read \\ \hline
% E1 & File Write \\ \hline
% E2 & Fork \\ \hline
% E3 & Execute \\ \hline
% E4 & LoadLibrary \\ \hline
% E5 & File Delete \\ \hline
% E6 & File Rename \\ \hline
% E7 & File Create \\ \hline
% E8 & File Property \\ \hline
% E9 & Exit \\ \hline
% E10 & LoadElf \\ \hline
% E11 & File Open \\ \hline
% E12 & File Close \\ \hline 
% E13 & Fork with shared open file \\ \hline
% E14 & File Open with close-on-exec mark \\ \hline
% E15 & File Mmap \\ \hline
% \end{tabular}
% \end{table}

\subsubsection{Transfer Rules}

To gather the semantics of the complete APT attack chain into target entity, we design a label transfer rule based on the ATT\&CK model, as shown in Table~\ref{tab:Transfer_Rules}. We divide APT attacks into five major stages, which are initial access, untrusted execution, lateral movement, suspicious behavior, and data exfiltration. Furthermore, the stage of suspicious behavior contains persistent stronghold, privilege escalation, credential access, and information collection. In Table~\ref{tab:Transfer_Rules}, label 1 and label 2 represent semantic labels carried by system entities, and the direction represents the flow direction of system events, which is also the label transmission direction. For example, if a process with label $PS1$ writes a file, we label the file with $FU2$, when a file with label $FU2$ is accessed by a process, we label the process with $PS3$. A detailed description of the transfer rules is in Appendix~\ref{exp:Transfer_Rules}. If a certain process contains some labels in above five stages, the process has the possibility to represent an ongoing APT attack, the APT judge rules will be described in Section~\ref{subsec:judgement_rules}.  

From the ATT\&CK model and TTP, it is known that each stage of the APT attack represents a tactic, and a tactic (i.e., initial access, untrusted execution, etc.) contains different techniques. Analysts can expand Table~\ref{tab:Transfer_Rules} according to different techniques under specific tactics.

\begin{table*}[h]\Large
\centering
\caption{Transfer rules. Label 1 and label 2 represent semantic labels carried by system entities, and the direction represents the flow direction of system events. D stands for front to back, R stands for back to front.}
\label{tab:Transfer_Rules}
\resizebox{\textwidth}{!}{
\begin{tabular}{|c|c|c|c|c|c|c|}
\hline
\multicolumn{2}{|c|}{\textbf{Stage}} & \textbf{Label1} & \textbf{Event} & \textbf{Label2} & \textbf{Direction} & \textbf{Description} \\ \hline
\multicolumn{2}{|c|}{\multirow{5}{*}{Initial Access}} & $PS1$ & Write & $FU2$ & D & A network-connected process wrote a file \\ \cline{3-7}
\multicolumn{2}{|c|}{~} & $PS3$ & Read & $FU2$ & R & The process reads a file containing network data \\ \cline{3-7}
\multicolumn{2}{|c|}{~} & $PS3$ & Write & $FU2$ & D & The process, which has accessed network data, writes these data to files \\ \cline{3-7}
\multicolumn{2}{|c|}{~} & $PS4$ & Read/Mmap & $FU1$ & R & The file uploaded by the user is loaded or read by processes \\ \cline{3-7}
\multicolumn{2}{|c|}{~} & $PS5$ & Read/LoadElf/Mmap & $FU3$ & R & The process interacts with non-existent files \\ \hline
\multicolumn{2}{|c|}{\multirow{3}{*}{Untrusted Execution}} & $PB1$ & Execute & $FU2$ & R & The file, which contains the codes from the network, is executed \\ \cline{3-7}
\multicolumn{2}{|c|}{~} & $PB1$ & LoadElf& $FU2$ & R & The file, which contains the codes from the network, is loaded \\ \cline{3-7}
\multicolumn{2}{|c|}{~} & $PB1$ & Write & $FU2$ & D & The process, which has executed the codes from the network, writes files \\ \hline
\multicolumn{2}{|c|}{\multirow{6}{*}{Lateral Movement}} & $PB4$ & Write & $FU4$ & D & The process, which has caused the Webshell attack, writes files \\ \cline{3-7}
\multicolumn{2}{|c|}{~} & $PB4$ & Read/Mmap & $FU4$ & R & The process reads or loads files written by Webshell \\ \cline{3-7}
\multicolumn{2}{|c|}{~} & $PB1$ & Write & $FU5$ & D & The process, which has caused the RAT attack, writes files \\ \cline{3-7}
\multicolumn{2}{|c|}{~} & $PB1$ & Read & $FU5$ & R & The process reads files written by RAT \\ \cline{3-7}
\multicolumn{2}{|c|}{~} & $PS5$ & Write & $FU6$ & D & The process, which has caused the Living-off-the-land attack, writes files \\ \cline{3-7}
\multicolumn{2}{|c|}{~} & $PS5$ & Read & $FU6$ & R & The process reads files written by Living-off-the-land  \\ \hline
\multirow{4}{*}{Suspicious Behavior} & Persistent Stronghold & $PB6$ & Write & $FH1$ & R & The process writes files that can control scheduled tasks such as /etc/crontab \\ \cline{2-7}
~ & Privilege Escalation & $PB7$ & Write & $FH2$ & R & The process writes files that can control permissions such as /etc/sudoers \\ \cline{2-7} 
~ & Credential Access & $PS6$ & Read & $FH3$ & R & The process reads files that hold sensitive information such as /etc/passwd \\ \cline{2-7} 
~ & Information Collection & $PS7$ & Read & $FH4$ & R & The process reads files that save historical commands such as .bash\_history \\ \hline
\multicolumn{2}{|c|}{\multirow{2}{*}{Data Exfiltration}} & $PB6-7|PS6-7$ & Write & $FH5$ & D & The process writes high-value data to files \\ \cline{3-7}
\multicolumn{2}{|c|}{~} & $PB8$ & Read & $FH5$ & R & The process reads files that contain high-value data \\ \hline
\end{tabular}}
\end{table*} 

\subsubsection{Judgement Rules}\label{subsec:judgement_rules}

% The judgment of the threat is carried out on the process node. Based on the label contained on the process node, we can roughly deduce what the process did. Finally, it can be judged whether a threat has occurred based on the different labels contained in the process, such as a process executes a file from network, and the process executes $sh$ or $bash$ command, then it can be concluded that the process may pose a RAT threat.

The attack judgment rules used in this paper are listed in Table~\ref{tab:judge_rules}. The last column specifies the prerequisites for the transfer rule to match. The prerequisites can specify conditions on the parameters of the alarm being matched. The administrator is able to judge whether an attack has occurred based on the different labels contained in the process. In other words, the labels of different entities on the attack chain will eventually come to a process through transfer and aggregation, when labels contained in the process match target judgment rules, the corresponding alerts (threats or APTs) will be reported to the administrator. For example, a process that contains the Webshell/RAT/Living-off-the-land attack labels, as well as the label $PB8$ may be the alert point of APT attack. A detailed description of the judgement rules is in Appendix~\ref{exp:judge_rules}.

\begin{table*}[h]\Large
\centering
\caption{Judgement rules for different attacks (threats and APTs). The prerequisites can specify conditions on the parameters of the alarm being matched. }
\label{tab:judge_rules}
\resizebox{\textwidth}{!}{
\begin{tabular}{|c|c|c|}
\hline
\textbf{Alert} & \textbf{Condition} & \textbf{Prerequisites} \\ \hline
\multirow{4}{*}{Download\&Execution} & \multirow{4}{*}{$PB1$} & $init(p) = match(p.name,"/(scp|wget)/") \rightarrow p.labels.add("PS1") $\\ 
~ & ~ & $propRd(p,f) = if(p.labels~contains~"PS1"~and~e == Write) \rightarrow f.labels.add("FU2") $ \\
~ & ~ & $propRd(f,p) = if(f.labels~contains~"FU2"~and~e == Execute) \rightarrow p.labels.add("PB1") $ \\
~ & ~ & $alert(p) = if(p.labels~contains~"PB1") \rightarrow Alert(Download\&Execution) $ \\\hline
\multirow{4}{*}{Webshell} & \multirow{4}{*}{$PS4$\&$PB4$} & $init(f) = match(f.name,"/(upload)/") \rightarrow p.labels.add("FU1") $\\ 
~ & ~ & $propRd(f,p) = if(f.labels~contains~"FU1"~and~e == Read or Mmap) \rightarrow p.labels.add("PS4") $ \\
~ & ~ & $inti(p) = match(p.name,"*") \rightarrow p.labels.add("PB4") $ \\
~ & ~ & $alert(p) = if(p.labels~contains~"PS4\&PB4") \rightarrow Alert(Webshell) $ \\ \hline
\multirow{3}{*}{RAT} & \multirow{3}{*}{$PB1$\&$PB5$} & $init(p) = match(p.name,"/(sh|bash)/") \rightarrow p.labels.add("PB5") $\\ 
~ & ~ & $inti(p) = if(p~cause~alert(Download\&Execution)) \rightarrow p.labels.add("FU2") $ \\
~ & ~ & $alert(p) = if(p.labels~contains~"PB1\&PB5") \rightarrow Alert(RAT) $ \\ \hline
\multirow{4}{*}{Living-off-the-land} & \multirow{4}{*}{$PS5$\&$PB5$} & $init(f) = match(f.name,"/(null)/") \rightarrow f.labels.add("FU3") $\\ 
~ & ~ & $propRd(f,p) = if(f.labels~contains~"FU3"~and~e == Load or Read) \rightarrow p.labels.add("PS5") $ \\
~ & ~ & $init(p) = match(p.name,"/(sh|bash)/") \rightarrow p.labels.add("PB5") $\\ 
~ & ~ & $alert(p) = if(p.labels~contains~"PS5\&PB5") \rightarrow Alert(Living-off-the-land) $ \\ \hline
\multirow{9}{*}{Suspicious Behavior} & \multirow{9}{*}{$PB6|PB7|PS6|PS7$} & $init(f) = match(f.name,"/etc/crontab") \rightarrow f.labels.add("FH1") $\\
~ & ~ & $propRd(f,p) = if(f.labels~contains~"FH1"~and~e == Write) \rightarrow p.labels.add("PB6") $\\ 
~ & ~ & $init(f) = match(f.name,"/etc/sudoers") \rightarrow f.labels.add("FH2") $ \\
~ & ~ & $propRd(f,p) = if(f.labels~contains~"FH2"~and~e == Write) \rightarrow p.labels.add("PB7") $\\
~ & ~ & $init(f) = match(f.name,"/etc/passwd") \rightarrow f.labels.add("FH3") $ \\
~ & ~ & $propRd(f,p) = if(f.labels~contains~"FH2"~and~e == Read) \rightarrow p.labels.add("PS6") $\\ 
~ & ~ & $init(f) = match(f.name,".bash\_history") \rightarrow f.labels.add("FH4") $ \\
~ & ~ & $propRd(f,p) = if(f.labels~contains~"FH2"~and~e == Read) \rightarrow p.labels.add("PS7") $\\ 
~ & ~ & $alert(p) = if(p.labels~contains~"PB6|PB7|PS6|PS7") \rightarrow Alert(Suspicious Behavior) $ \\ \hline
\multirow{4}{*}{Data Exfiltration} & \multirow{4}{*}{$PB8$} & $init(p) = if(p~cause~alert(Suspecious Behavior)) \rightarrow p.labels.add("PB6|PB7|PS6|PS7") $\\ 
~ & ~ & $propRd(p,f) = if(p.labels~contains~"PB6|PB7|PS6|PS7"~and~e == Write) \rightarrow f.labels.add("FH5") $ \\
~ & ~ & $propRd(f,p) = if(f.labels~contains~"FH5"~and~e == Read) \rightarrow p.labels.add("PB8") $\\ 
~ & ~ & $alert(p) = if(p.labels~contains~"PB8") \rightarrow Alert(Data Exfiltration) $ \\ \hline
APT & $((PS4\&PB4)|(PB1\&PB5)|(PS5\&PB5))\&PB8$ & $alert(p) = if(p~cause~alert(Webshell|RAT|Living-off-the-land)~and~p.labels~contains~"PB8") \rightarrow Alert(APT) $ \\ \hline
\end{tabular}}
\end{table*}

\section{Design Experiments}
\label{sec:design_experiments}

In our evaluation, we first describe the experimental setup. Then we introduce the effectiveness of data compaction, the accuracy and effectiveness of APT detection, and the overhead of the system in turn. 

\subsection{Experimental Setup}

Our datasets consist of two parts, one is collected from our laboratory, and the other is from Darpa Engagement. Table~\ref{tab:datasets} summarizes the property of our dataset.  

\begin{table*}[h]%
\centering
\caption{Details of our Datasets. L stands for data from laboratory, and E stands for data from Darpa Engagement. }
\label{tab:datasets}
\begin{tabular}{|c|c|c|c|c|}
\hline
\textbf{Dataset} & \textbf{Duration (hours)} & \textbf{Platform}     & \textbf{Source}  & \textbf{Attack Description}                                                                                          \\ \hline
L-1              & 24                        & Ubuntu 16.04 (64 bit) & Laboratory       & Webshell attack from the backdoored Apache                                                                            \\ \hline
L-2              & 24                        & Ubuntu 16.04 (64 bit) & Laboratory       & Remote Access Trojan from the phishing website                                                                       \\ \hline
L-3              & 24                        & Ubuntu 16.04 (64 bit) & Laboratory       & Living-off-the-land attack from vulnerable service                                                                   \\ \hline
E-1              & 168                       & Ubuntu 14.04 (64 bit) & Darpa Engagement & \begin{tabular}[c]{@{}c@{}}Information gather and exfiltration \&\\ Malicious file download and execute\end{tabular} \\ \hline
E-2              & 168                       & Ubuntu 14.04 (64 bit) & Darpa Engagement  & In-memory attack with firefox                                                                                        \\ \hline
\end{tabular}
\end{table*}

% \begin{table}[h]\tiny%
% \centering
% \caption{Datasets and scenes.}
% \label{tab:datasets}
% \resizebox{\columnwidth}{10mm}{
% \begin{tabular}{|c|c|c|c|c|}
% \hline
% \textbf{Stream No.} & \textbf{Duration} & \textbf{Platform} & \textbf{Scenario Name} & \textbf{Attack Surface}\\ \hline
% L1 & 24h & Ubuntu 16.04(64bit) & Webshell & Backdoored Apache \\ \hline
% L2 & 24h & Ubuntu 16.04(64bit) & Remote Access Trojan & Backdoor \\ \hline
% L3 & 24h & Ubuntu 16.04(64bit) & Living off the Land & Vulnerable Procedure \\ \hline
% E1-2 & 168h & Ubuntu 14.04(64bit) & Dataset-A from Darpa Engagement & Vulnerabilities \\ \hline
% E3 & 168h & Ubuntu 14.04(64bit) & Dataset-B from Darpa Engagement & In-memory attack \\ \hline
% \end{tabular}}
% \end{table}

\subsubsection{The Dataset from Laboratory}

For the data from our laboratory, two participants (red team) in our laboratory were responsible for instrumenting OS and carrying out attack campaigns (on three hosts with Ubuntu 16.04), while the other two participants (blue team) performed data collection, data compaction and attack detection in a real-time manner. The benign background activities were also being carried out on the machines used for experimentation, such as web browsing, chatting, and document editing. In order to get an adversarial engagement, the blue team had no prior knowledge of the attacks planned by the red team. As shown in Table~\ref{tab:datasets}, L-1, L-2, and L-3 were the APT datasets generated by simulating real scenes in our laboratory, and the collection of each dataset lasted for 24 hours. Each dataset contained normal data and attack data. The attacks contained in above three datasets (L-1, L-2, and L-3) are: Webshell attack from the backdoored Apache, Remote Access Trojan from the phishing website, and Living-off-the-land attack from vulnerable service. A detailed description of attacks contained in L-1, L-2, and L-3 are in Appendix~\ref{exp:lab_attacks}.

\subsubsection{The Dataset from Darpa Engagement}

For the data from Darpa Engagement, two hosts with Ubuntu 14.04 were deployed in advance for the data collection (blue team) and the  APT attack (red team). Note that Blue team had no prior knowledge about the attack prepared by the red team. Similarity, when the red team was attacking the host, other activities of normal programs on the host were also proceeding simultaneously. Benign activities contained browsing websites, downloading and executing binary files, reading/writing emails and documents, etc.. Overall, more than 99\% of the events in the dataset had nothing to do with attacks. As shown in Table~\ref{tab:datasets}, E-1 and E-2 are the APT datasets generated by Darpa Engagement, and the collection of each dataset lasted for 168 hours. Each dataset contained normal data and attack data. The attacks contained in above two datasets (E-1 and E-2) are: Information gather and exfiltration, Malicious file download and execute, and In-memory attack with firefox. A detailed description of attacks contained in E-1 and E-2 are in Appendix~\ref{exp:darpa_attacks}.

\subsection{System Performance}

Here, we will introduce the effectiveness of data compaction, the accuracy and effectiveness of APT detection, and the system overhead.

\subsubsection{Performance of Data Compaction}

\textbf{(1) Effectiveness of Redundant Semantics Skipping}

In real-time APT detection, the strategy of redundant semantics skipping is adopted to reduce events with repetitive semantics. It can be carried out in real-time streaming data to ensure the performance of the detection system (i.e., reduce memory overhead and running time). Table~\ref{tab:redundant_semantics_skipping} gives results of redundant semantics skipping.

\begin{table}[h]
\centering
\caption{Results of Redundant Semantics Skipping. }
\label{tab:redundant_semantics_skipping}
\resizebox{\columnwidth}{!}{
\begin{tabular}{|c|c|c|c|}
\hline
\textbf{Dataset} & \textbf{Sum of Events} & \textbf{Skip of Events} & \textbf{Saved Time (ms)/1000K} \\ \hline
L-1 & 1142480 & 381885 & 1716 \\ \hline
L-2 & 1070143 & 357307 & 1038 \\ \hline
L-3 & 1006603 & 335721 & 2514 \\ \hline
E-1 & 27978043 & 6019924 & 7790 \\ \hline
E-2 & 33743425 & 6016741 & 11050 \\ \hline
\end{tabular}}
\end{table}

For the datasets (L-1, L-2, and L-3) from our laboratory, each dataset contains about 1000K events, and about 350K pieces of events can be skipped on average through the redundant semantic skipping strategy. In the real-time APT detection, \sn{} can save 1000ms-2500ms for every 1000K samples. For the datasets (E-1 and E-2) from Darpa Engagement, the datasets contain about 28000K-34000K events, and about 6000K pieces of events are able to be skipped on average through the redundant semantic skipping strategy. Although the proportion of skipped events has decreased compared with the data from laboratory (down from 35\% to 20\%), the time saved per 1000K events has increased significantly, reaching 7800ms-11000ms.

We find that \sn{} can save more time on the Darpa Engagement dataset, the reason is that the process of file reading and writing events takes more time, and there are a large number of such events on the Darpa Engagement dataset.

\textbf{(2) Effectiveness of Non-viable Entity Pruning}

As the running time increases, incremental entities will be saved. However, some non-viable entities will neither pose a threat nor affect the forensic analysis. Therefore, we can delete these non-viable entities based on EXIT events and PHF labels to reduce memory consumption. Results of non-viable entities pruning are shown in Table~\ref{tab:non-viable_entities_pruning}. The results on different datasets show that with the growth of time and the increase of entities, the proportion of pruned processes is about 3\% to 8\%. Moreover, some inactive files will be moved to the disk (inactivity is defined as no change for more than 5 minutes) for reducing overhead. We can find from Table~\ref{tab:non-viable_entities_pruning} that the pruned files mainly depend on the degree of interaction between files and processes in the dataset. The reason file pruning ratio of L-3 and E-1 is relatively high is because these two datasets contain a large number of inactive temporary files. These temporary files are all moved to disk. In addition, we also evaluate the CPU and memory overhead of \sn{} with/without pruning, which will be discussed in detail in Section~\ref{subsec:overhead}. 

\begin{table}[h]
\centering
\caption{Results of Non-viable Entities Pruning. Sum stands for the total number of entities (processes or files). Remain stands for the number of entities (processes or files) remaining after being pruned.}
\label{tab:non-viable_entities_pruning}
\resizebox{\columnwidth}{!}{
\begin{tabular}{|c|c|c|c|c|c|c|}
\hline
\multirow{2}{*}{\textbf{Dataset}} & \multicolumn{3}{c|}{\textbf{Prune for Process}}                                                   & \multicolumn{3}{c|}{\textbf{Prune for File}}                                                      \\ \cline{2-7} 
                                  & \textbf{Sum} & \textbf{Remain} & \textbf{\begin{tabular}[c]{@{}c@{}}Pruning\\ Ratio (\%)\end{tabular}} & \textbf{Sum} & \textbf{Remain} & \textbf{\begin{tabular}[c]{@{}c@{}}Pruning\\ Ratio (\%)\end{tabular}} \\ \hline
L-1                               & 125          & 121             & 3.2                                                              & 144          & 138             & 4.2                                                              \\ \hline
L-2                               & 91           & 84              & 7.7                                                              & 255          & 247             & 3.1                                                              \\ \hline
L-3                               & 127          & 119             & 6.3                                                              & 138          & 53              & 61.6                                                             \\ \hline
E-1                               & 16926        & 16226           & 4.1                                                              & 1360         & 114             & 91.6                                                             \\ \hline
E-2                               & 21673        & 20829           & 3.9                                                              & 1232         & 1209            & 1.9                                                              \\ \hline
\end{tabular}}
\end{table}

% The memory consumption is shown in the Figure \ref{fig:memoryconsumption} below:

% \begin{figure}[h]
%     \centering
%     \includegraphics[width=0.9\columnwidth]{fig/memoryconsumption.png}
%     \caption{Memory Consumption}
%     \label{fig:memoryconsumption}
% \end{figure}

% The figure shows that the memory after pruning consume more than the before at the beginning of the program. The reason for this phenomenon may be that the pruning operation we made to the program caused a certain amount of memory consumption. But in the later stage of the program (shown in the figure after 360s), the memory before pruning will consume more than the after.

\textbf{(3) Comparison with Related Work}

In \sn{}, we use a data compaction method that maintains global semantics to delete the equivalent information flow to the same target entity with low overhead by using the semantic attributes of the source entity. In addition, our APT detection and data compaction can be performed at the same time to ensure real-time performance. As shown in Table~\ref{tab:comparison_compaction}, we compared \sn{} to other existing data compaction methods from the following three aspects: real-time performance, efficiency, and overhead.

LCD \cite{xu2016high} considers the event compaction of the single node in a real-time manner, but it does not link the contextual semantics, which could cause difficulty in forensics analysis. Moreover, LCD cannot handle network events. The input of FD \cite{hossain2018dependence} is the whole dependence graph, which can only start after data collection is completed (i.e., FD uses offline cached data). Also, FD cannot deal with events of subprocess/thread. GS \cite{zhu2021general} is able to establish global dependencies as FD does, but the compaction system of GS does not have a discarding mechanism (e.g., the processing of lengthy list fields, killed processes, and unused files). As time increases, the real-time performance and efficiency of the system will decrease. Our compaction method is able to improve the efficiency of real-time APT detection and reduce data  storage overhead for forensic analysis by adopting redundant semantics skipping and non-viable entity pruning. All in all, the requirements of real-time performance, efficiency, and overhead can be jointly addressed by \sn{}. It outperforms all the above related methods.

% LCD can only consider the characteristics of a single node, but not the global semantics. Only when there is no outgoing edge between the outgoing edge and the outgoing edge of the vertex V and no outgoing edge between the incoming edge, the LCD allows the reduction under two strict conditions. Since each outgoing edge of the file has a corresponding incoming edge, the condition of the merged edge implemented in the LCD cannot be met, all events will be retained, and it cannot handle network events.

% Compare the LCD algorithm with our compression algorithm. The results are shown in the Table. Compared with the LCD algorithm, our compression algorithm is slightly optimized than the LCD algorithm. The reason is that we have considered the network connection that the LCD algorithm does not take into account.

\begin{table}
\centering
\caption{Comparison with related work. L represents low, M represents medium, and H represents high.}
\label{tab:comparison_compaction}
\resizebox{\columnwidth}{!}{
\begin{tabular}{|c|c|c|c|}
\hline
\textbf{Method} & \textbf{Real-time} & \textbf{Efficiency} & \textbf{Overhead} \\ \hline
\textbf{\sn{}}   & H                  & H                   & L                 \\ \hline
\textbf{LCD (2016) \cite{xu2016high}}    & H                  & L                   & L                 \\ \hline
\textbf{FD (2018) \cite{hossain2018dependence}}     & L                  & M                   & H                 \\ \hline
\textbf{GS (2021) \cite{zhu2021general}}     & M                  & H                   & H                 \\ \hline
\end{tabular}}
\end{table}

% \begin{table}
%     \centering
%     \caption{Comparison to LCD Algorithm.}
%     \label{tab:comparison_to_LCD}
%     \begin{tabular}{|l|l|l|l|l|l|}
%     \hline
%          & L1 & L2 & L3 & E1 & E2 \\ \hline
%         LCD & 66.71\% & 66.72\% & 66.82\% & 78.73\% & 82.45\% \\ \hline
%         \tt{Our} & 66.57\% & 66.61\% & 66.65\% & 78.48\% & 82.17\% \\ \hline
%     \end{tabular}
% \end{table}

% The FD algorithm establishes global dependencies to reduce redundant events on offline cached data. It is precisely because of this characteristic of the FD algorithm that the real-time performance of the algorithm is very poor. In terms of compression coverage, the FD algorithm does not compress subprocess/thread which are in our algorithm.

% The GS algorithm is similar to the FD algorithm and is also establishes global dependencies. So the real-time performance of our algorithm is better than them. Our algorithm adds pruning to non-viable nodes based on the implementation of the GS algorithm. Reduce the running overhead during real-time detection, and delete the events of these non-viable nodes during compression.

\subsubsection{Accuracy and Effectiveness of APT Detection}
In this section, we will explain how the APT detection framework can effectively detect attacks in Table~\ref{tab:datasets}.

\begin{figure}[h]
    \centering
    \includegraphics[width=0.9\columnwidth]{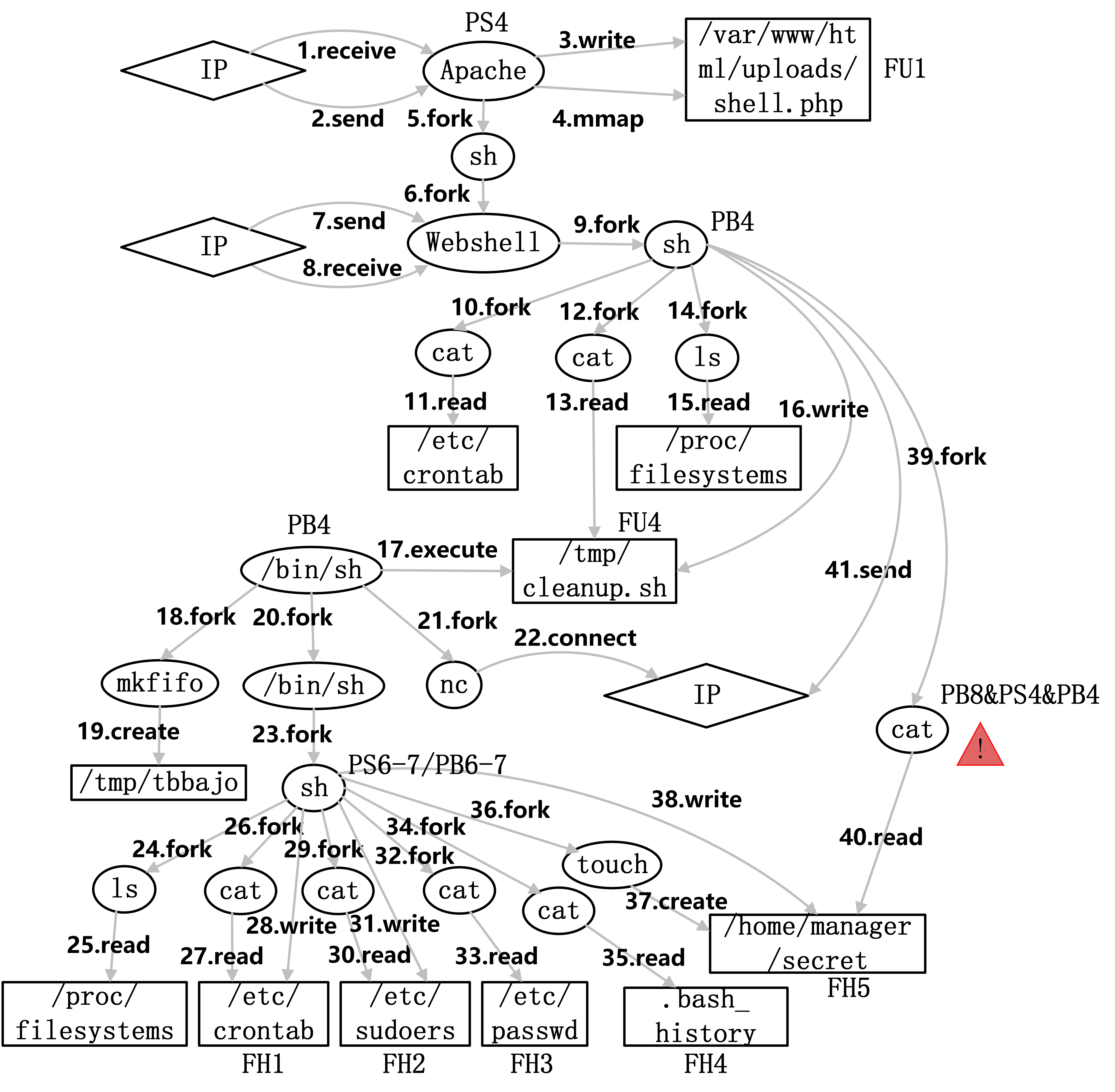}
    \caption{APT attack chain that uses Webshell as the entrance.}
    \label{fig:webshell}
\end{figure}

\textbf{L-1: Webshell attack from the backdoored Apache}. The APT attack chain in dataset L-1 is shown in Figure~\ref{fig:webshell}. The process of the APT detection framework is as follows: 

(1) The folder directory $/var/www/html/uploads/$ in the web service is to save uploaded files. When the Webshell file $shell.php$ has been uploaded to this directory by the Apache vulnerability, we lable this file with $FU1$ according to the file label definition (time point 1-3). 

(2) If the file $shell.php$ is accessed or loaded by a process, the process will be labeled with $PS4$ by the transfer rule $(PS4, READ/MMAP, FU1, R)$ (time point 4). Note that the Apache in Figure~\ref{fig:webshell} is a subprocess, which will not cause label taint (i.e., other behaviors of apache will not be labeled). 

(3) When a process with $PS4$ executes any commands, the process will be labeled with $PB4$ (time point 5-6). In order to track the spread of attacks after the invasion, \sn{} will label files accessed by the process (with $PB4$) as $FU4$ according to the transfer rule $(PB4, WRITE, FU4, D)$, which indicates that these files may be threatened by the subsequent proliferation of Webshell. When the file is accessed by other processes, the semantics will be passed back to the process (labeled with $PB4$) by the transfer rule $(PB4, READ/MMAP, FU4, R)$, which makes subsequent infiltrations of Webshell fully recorded and detected. 

(4) The process of Webshell ($PB4$) writes a controllable script $(/tmp/cleanup.sh)$, $(/tmp/cleanup.sh)$ will be labeled with $FU4$ by the transfer rule $(PB4, WRITE, FU4, D)$ (time point 9 and 16). 

(5) The script ($FU4$) is loaded and executed by a privileged process $(/bin/sh)$, $(/bin/sh)$ will be labeled with $PB4$ by the transfer rule $(PB4, READ/MMAP, FU4, R)$, $(/bin/sh)$ executes $nc$ to create command\&control between the user and the attacker (time point 17-23).

(6) The process $sh$ ($PB4$) continues to access sensitive files, such as $/etc/crontab (FH1)$, $/etc/sudoers (FH2)$, $/etc/passwd (FH3)$, and $.bash\_history (FH4)$, it will be labeled with the corresponding label (i.e., $PB6$, $PB7$, $PS6$, and $PS7$) indicating that it has read or written these sensitive files (time point 26-35). 

(7) When a process $sh$ has read or written sensitive files that wants to pass high-value information out, the file $secret$ written by it will be labeled with $FH5$ (time point 36-38), indicating that the file $secret$ may contain high-value information $(PS6-7/PB6-7, WRITE, FH5, D)$. 

(8) If another process $cat$ subsequently accesses the file with the $FH5$ label, it may be labeled with $PB8$ due to the threat of data exfiltration $(PB8,READ,FH5,R)$ (time point 39-40). Finally, the process that satisfies this series of attack chains may generate the threat of APT attacks (including Initial Access, Lateral Movement, Suspicious Behavior, and Data Exfiltration). Therefore, it is reported that there may be APT which uses Webshell as the entrance to the process that contains labels of Webshell and $PB8$ at the same time.

\textbf{L-2: Remote Access Trojan from the phishing website}. For the APT in L-2, the detection flow of \sn{} is as follows:

(1) Through a phishing link, the target host downloads the Trojan file $(vpn.elf)$. Since the file is transmitted over the Internet, it will be labeled with $FU2$. 

(2) After a period of time, the file $(vpn.elf)$ is executed, and the target process will be labeled with $PS1$ by the transfer rule $(PS1, EXECUTE, FU2, R)$, indicating that the process executes files originating from the Internet may cause a threat. 

(3) Subsequent attacks and proliferation are the same as they are in the Webshell experiment. Finally, it is reported that there may be APT which uses RAT as the entrance for the process that contains labels of RAT and $PB8$ at the same time.

\textbf{L-3: Living-off-the-land attack from vulnerable service}. For the APT in L-3, the detection flow of \sn{} is as follows:

(1) Through an exploit script, the threatening process will interact with a file named $"(null)"$. Therefore, the non-existent file will be labeled with $FU3$.

% We have run the program with heap/stack overflow on the target machine in advance, and have an exploit script to obtain the shell. When we know that the target host is running a program with heap/stack overflow on port 29273, we can use this script to obtain permission. By observing the collected datasets, it is found that the threatening process will interact with a file named $"(null)"$. Therefore, we label the non-existent file $(FU3)$. 

(2) When a process interacts with the file, the process will be labeled with $PS5$ by the transfer rule $(PS5, READ/MMAP/LOADELF, FU3, R)$. When the process containing the label $PS5$ continues to execute the $sh$ or $bash$ command, it will be labeled with $PB5$, which means that the process has got the shell and carried out a in-memory attack. 

(3) Subsequent attacks and proliferations are the same as they are in the Webshell experiment. Finally, it is reported that there may be APT that uses Living-off-the-land as the entrance for the process that contains labels of Living-off-the-land and $PB8$ at the same time.

\textbf{E-1: Information gather and exfiltration}. The detection flow of \sn{} is as follows:

(1) The file ($/etc/passwd$) containing sensitive information are labeled with $FH3$. 

(2) When a process reads this file, the process will be labeled with $PS6$ by the transfer rule $(PS6, READ, FH3, R)$. 

(3) The process writes the high-value data on $/dev/pts/1$, $/dev/pts/1$ will be labeled with $FH5$ by the transfer rule $(PS6, WRITE, FH5, D)$. By this way, the user may have exfiltrated this information.

\begin{figure}[h]
    \centering
    \includegraphics[width=0.9\columnwidth]{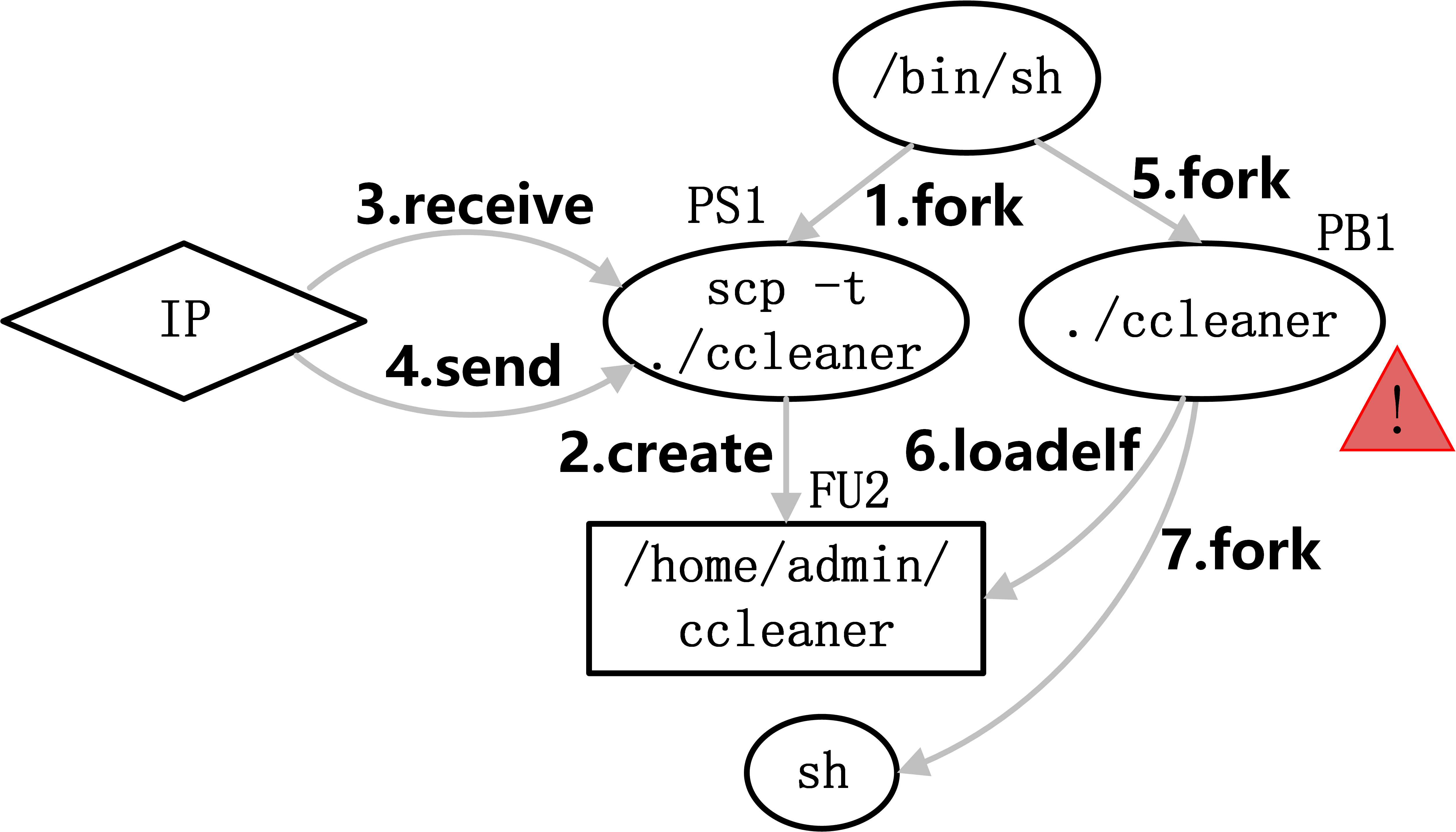}
    \caption{APT attack chain that uses ccleaner as the entrance.}
    \label{fig:ccleaner}
\end{figure}

\textbf{E-1: Malicious file download and execute}. The attack chain is shown in Figure~\ref{fig:ccleaner}. The detection flow of \sn{} is as follows:

(1) A file named $ccleaner$ is downloaded to the local through $scp$, this file will be labeled with $FU2$. Because the process that uses the $scp$ command is likely to include an internet connection $PS1$, the file will be labeled with $FU2$ by the transfer rule $(PS1,WRITE,FU2,D)$. 

(2) After the file is executed by the process, the process will be labeled with $PB1$ by the transfer rule $(PB1, EXECUTE, FU2, R)$. Therefore, the process containing the $PB1$ label may pose a threat (download and execution).

\begin{figure}[h]
    \centering
    \includegraphics[width=0.9\columnwidth]{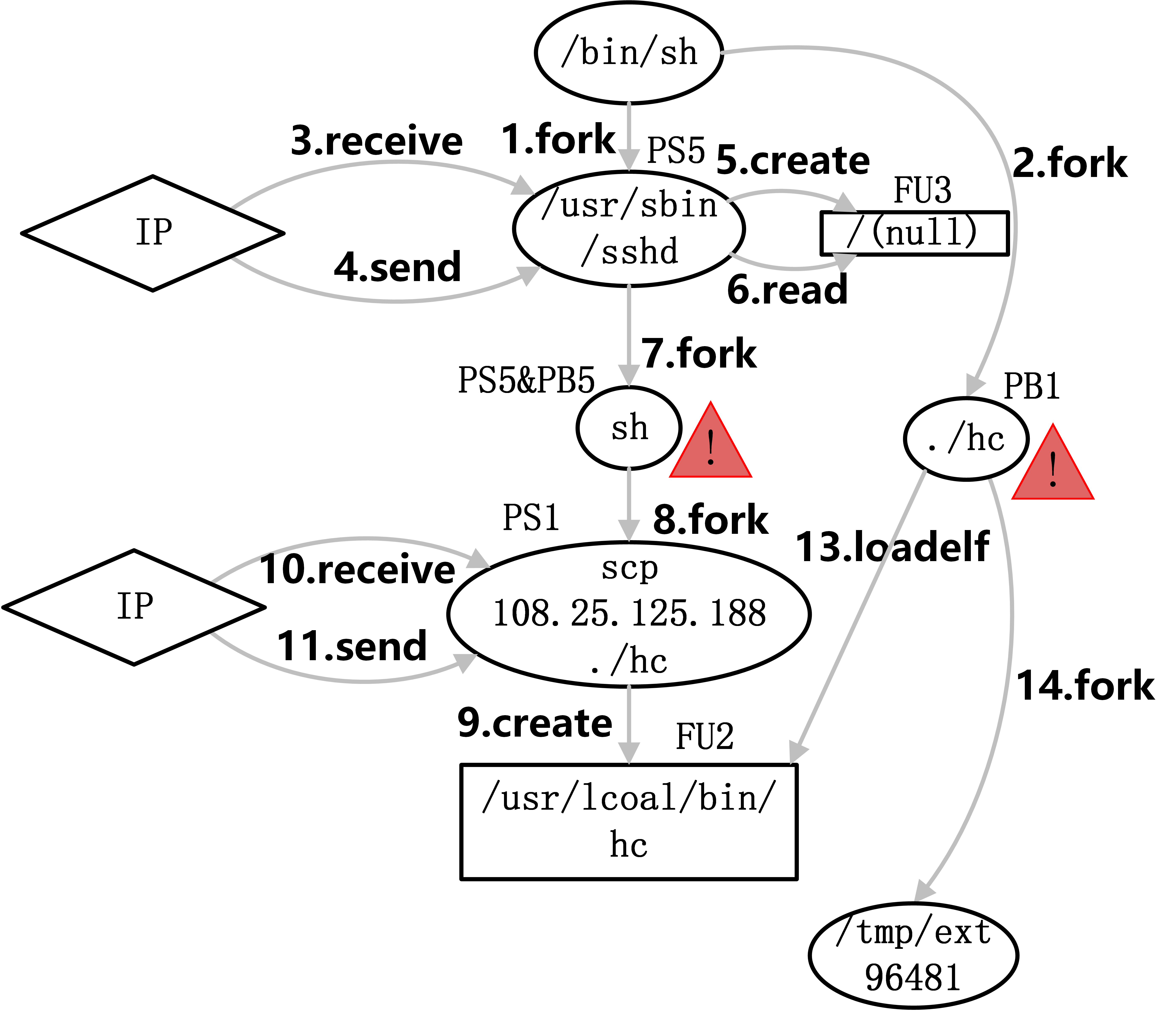}
    \caption{APT attack chain that uses in-memory attack as the entrance.}
    \label{fig:hc}
\end{figure}

\textbf{E-2: In-memory attack with firefox}. The attack chain is shown in Figure~\ref{fig:hc}. The detection flow of \sn{} is as follows:

(1) Use Firefox's write-executable memory space to cause in-memory attacks through $sshd$ process. The sshd process interacts with the $"/(null)"$ file, and the process is labeled with $PS5$ by the transfer rule $(PS5, READ/MMAP/LOADELF, FU3, R)$.

(2) Later, the process with $PS5$ gets the shell, it will be labeled with $(PB5)$. Processes that contain the $PB5$ and $PS5$ tags may cause the threat of in-memory attacks. 

(3) Finally, a file named $hc$ was uploaded and executed through the $sshd$ in-memory attack. The detection flow is the same as the $ccleaner$ attack.

\subsection{Overhead of \sn{}}\label{subsec:overhead}

\subsubsection{CPU and Memory Usage}

The CPU and memory usage of \sn{} for five different datasets are shown in Table~\ref{tab:usage_CPU_Mem}. We use performance monitor to record the average overhead of the system while it is running. Here, the CPU usage means the usage rate on a single-core CPU (thread). In the absence of pruning, the average single-core CPU usage of \sn{} is 5.3\%, 5.8\%, 4.7\%, 9.7\%, and 10.1\% on L-1, L-2, L-3, E-1, and E-2, respectively, while the average memory usage is 90.1 MB, 99.7 MB, 85.6 MB, 117.7 MB, and 130.3 MB on L-1, L-2, L-3, E-1, and E-2, respectively. By deploying non-viable entity pruning (to further reduce the system overhead), the average single-core CPU usage is reduced by 1.1\%, 0.7\%, 0.4\%, 2.0\%, and 1.6\% on L-1, L-2, L-3, E-1, and E-2, respectively, while the average memory usage is reduced by 14.8 MB, 22.6 MB, 14.9 MB, 22.2 MB, and 19.7 MB on L-1, L-2, L-3, E-1, and E-2, respectively. All in all, in the case of real-time detection, \sn{} can maintain low CPU usage and satisfactory memory overhead. 

\begin{table}[h]
    \centering
    \caption{The average overhead of \sn{}. By deploying non-viable entity pruning, the overheads of CPU and memory are reduced.}
    \label{tab:usage_CPU_Mem}
    \resizebox{\columnwidth}{!}{
    \begin{tabular}{|c|c|c|c|c|}
    \hline
    \multirow{2}{*}{\textbf{Dataset}} & \multicolumn{2}{c|}{\textbf{Without Pruning}} & \multicolumn{2}{c|}{\textbf{With Pruning}} \\ \cline{2-5} 
                                      & \textbf{CPU (\%)}    & \textbf{Memory (MB)}   & \textbf{CPU (\%)}  & \textbf{Memory (MB)}  \\ \hline
    L-1                                & 5.3                  & 90.1                     & 4.2\%              & 75.3                    \\ \hline
    L-2                                & 5.8                  & 99.7                     & 5.1\%              & 77.1                    \\ \hline
    L-3                                & 4.7                  & 85.6                     & 4.3\%              & 70.7                    \\ \hline
    E-1                                & 9.7                  & 117.7                    & 7.7\%              & 95.5                    \\ \hline
    E-2                                & 10.1                 & 130.3                    & 8.5\%              & 110.6                   \\ \hline
    \end{tabular}}
\end{table}

\subsubsection{Real-Time Performance}

\begin{figure}[h]
    \centering
    \includegraphics[width=0.9\columnwidth]{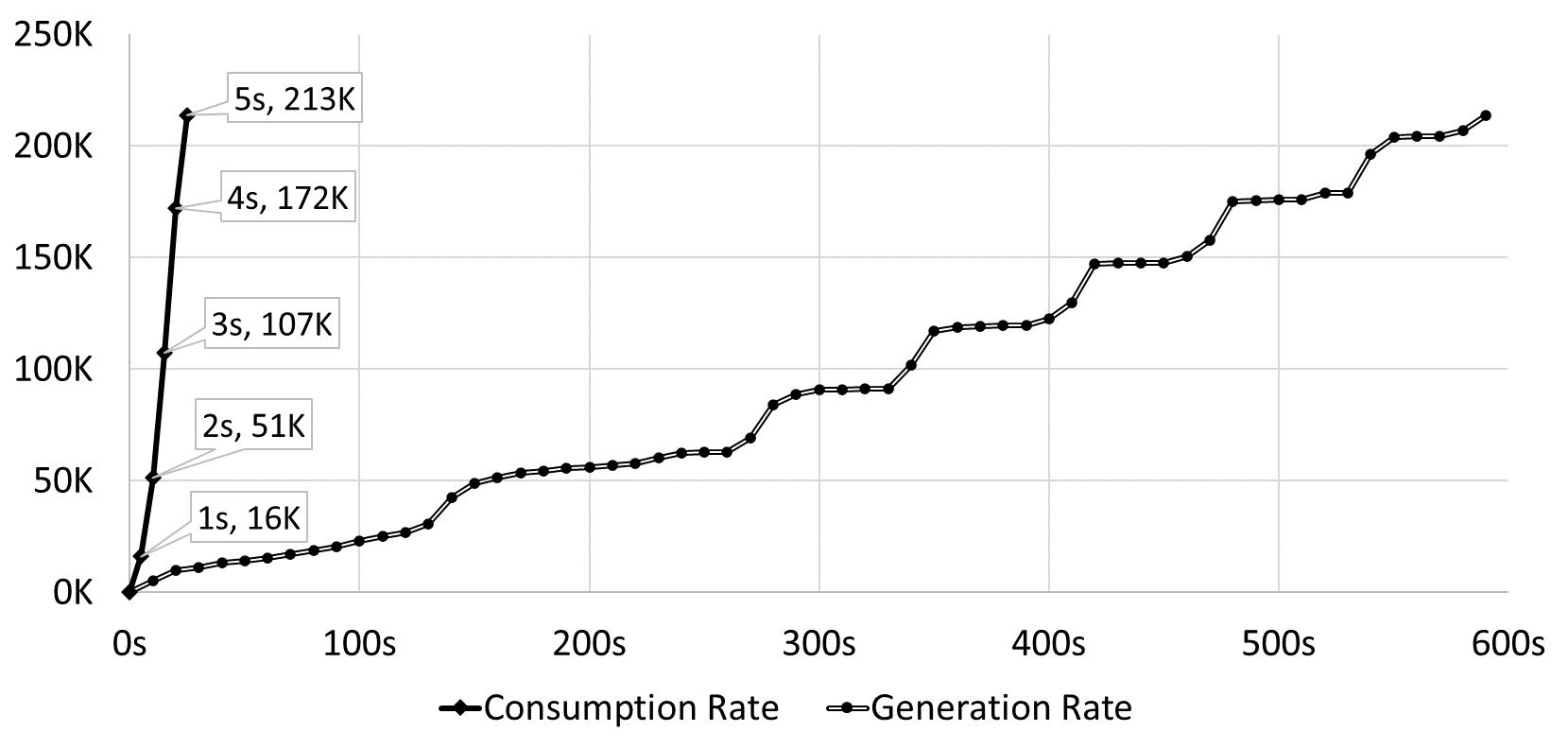}
    \caption{Real-time Performance of \sn{}. Comparison of data consumption rate and data generation rate.}
    \label{fig:realtime}
\end{figure}

We test the real-time performance of the \sn{} by comparing data consumption rate and data generation rate. In the experiment, the personnel in our laboratory performed normal operations on Linux host for 600 seconds, making the collected content as similar as possible to five datasets (L-1, L-2, L-3, E-1, and E-2). Results are shown in Figure~\ref{fig:realtime}. 

In Figure~\ref{fig:realtime}, the line on the right shows that about 230K pieces of data (contain entities and events) are generated and transmitted in real-time within ten minutes. The smooth phase of the line in Figure~\ref{fig:realtime} stands for the idle state of the host, while the steep phase of the line stands for the busy state of the host, and the transmission rate is about 400 per second. The other line represents the consumption rate of \sn{}. About 213K pieces of data can be processed within 5 seconds, and the average processing rate is about 42K per second. The consumption rate is far greater than the generation rate, which shows the real-time of \sn{} is feasible.

\subsubsection{Comparison with Related Work}

We compare \sn{} with state-of-the-art studies to show the advantage of our work.

\textbf{(1) Conan \cite{xiong2020conan}}. Conan only considers the data compaction in the storage stage, while \sn{} can perform compaction and detection at the same time. In addition, \sn{} also prunes invalid/unused entities to reduce overhead. In our experiment, the average memory usage of Conan on the dataset from laboratory (L-1, L-2, and L-3) is about 90 MB. For the dataset from Darpa Engagement, the average memory usage of Conan is about 140 MB.  Compared with \sn{}, the memory overhead of Conan is increased by about 30\% to 50\%, and the growth rate of CPU usage reaches to 10\% to 20\%. Furthermore, \sn{} has a better semantic interpretation and good detection efficiency for in-memory attacks.

\textbf{(2) Sleuth \cite{hossain2017sleuth} \& Holmes \cite{milajerdi2019holmes}}. According to the description in Section~\ref{sec:related}, Sleuth will read all the data into the memory at one time during detection, which cannot ensure real-time performance. The memory Sleuth used is related to the size of the dataset. In our experiment, the memory usage of Sleuth on L-1, L-2 and L-3 ranges from 80 MB to 160 MB, while on E-1 and E-2, it ranges from 300 MB to 400 MB. For Holmes, authors in \cite{milajerdi2019holmes} showed that a nearly linear growth in memory consumption. In our experiment, the memory usage of Holmes on L-1, L-2 and L-3 ranges from 90 MB to 150 MB, while on E-1 and E-2, it ranges from 500 MB to 1 GB. Compared with \sn{}, the memory overhead of Sleuth and Holmes is increased by at least 60\% and 20\%, respectively. \sn{} has a good real-time performance and can quickly respond to threats that occur in the system.

\section{Conclusion \& Discussion}
\label{sec:conclusion}

In this paper, we design a stable, efficient, and real-time APT detection system, called \sn{}. Unlike previous studies with the problems of the weak data source integrity, large data processing overhead and poor real-time performance, \sn{} can collect reliable, stable and semantically rich data sources via audit, reduce the overhead of the detection system in real-time via redundant semantics skipping and non-viable node pruning, and carry out real-time APT attack response and alarm through the transfer and aggregation of labels based on ATT\&CK model. Experimental results on both laboratory and Darpa Engagement show that \sn{} can effectively detect web vulnerability attacks, file-less attacks and remote access trojan attacks, at the same time has a low false positive rate, which adds far more value than the existing frontier work. And we will discuss the limitations and improvements of our work.

\subsection{False Positive}

Although \sn{} did not show any false positives in the experiment of this article, we still need to discuss the following scenarios:

\textbf{Normal Meta behavior}. Since an APT attack is composed of multiple different behaviors, each individual behavior (we call it meta behavior) does not represent the occurrence of an attack. For example, $READ$ events to high-value files such as $/etc/passwd$ may occur during the APT attack stage (when the user is authenticated, $/etc/passwd$ will be accessed), but related processes should not be identified as threats. In order to prevent single-point false positives, \sn{} adopts the method of contextual semantic transfer to detect the APT attack chain. Analysts can set stringent judgment rules to avoid false positives. 

\textbf{Taint explosion}. Recall that through the transfer and aggregation of labels, we are able to correlate the context information of the attack. However, this method is prone to cause taint explosion (e.g., polluting the daemon process of the system at the entry point, causing all entities interacting with the process to be polluted). A lot of false positives will occur due to taint explosion. In order to solve this problem, we can formulate stricter labeling and transfer rules to ensure that the daemon process is not easily marked as suspicious when it has not been invaded or modified. 

\subsection{False Negative}

Reducing the false negative rate is a major challenge for APT detection, because APT attacks are completely unknown. \sn{} uses different tactical stages in the ATT\&CK model to describe a complete APT attack. In APT detection, \sn{} can reflect the nature of the attack by tracking untrusted data flow, untrusted control flow, and high-value data flow. However, for some alternative attacks, such as open source software supply chain attacks (i.e., to inject malicious code into open source software packages in order to compromise dependent systems further down the chain) \cite{ohm2020backstabber}. Some well-known open source softwares may be whitelisted by system administrators, allowing them to be released at the system entry point. To solve this problem, we can combine with code dependency analysis at the entry point to label unsafe factors and reduce false negatives.

%%%%%%%%% REFERENCES
{\small
\bibliographystyle{ieee_fullname}
\bibliography{egbib}
}

%%%%%%%%% APPENDIX
\clearpage
\appendix
\setcounter{equation}{0}
\pagestyle{empty}

\subsection{Explanation of Defined Label in Table~\ref{tab:process_label}}
\label{exp:process_label}

$PS1$: If a process has network connection, we label it with $PS1$, for we can't trust the data from network.

$PS2$: If a process has accessed high-value data flow nodes, we label it with $PS2$, for the process holds sensitive data.

$PS3$: If a process has data from the network, we label it with $PS3$, e.g., the process reads the file downloaded from network.

$PS4$: If a process has loaded or read the uploaded file from network, which can cause Webshell attack, we label it with $PS4$ for the uploaded file can't be trusted.

$PS5$: If a process has interacted with non-existent files, which can cause Living-off-the-land attack, we label it with $PS5$ for accessing to non-existent files is a feature of Living-off-the-land attacks.

$PS6$: If a process has read the file, which contains sensitive information such as $/etc/passwd$, we label it $PS6$ for it may cause leakage of sensitive information.

$PS7$: If a process has read the file, which saves historical commands such as $.bash\_history$, we label it with $PS7$ for the traces may be exploited.

$PB1$: If a process has executed a file from network, we label the process with $PB1$, for the file can't be trusted.

$PB2$: If a process has executed a sensitive file, we label the process with $PB2$, for the file may cause attack.

$PB3$: If a process has executed a sensitive command, we label the process with $PB3$, for the command may cause attack.

$PB4$: If a process has executed any command, we label the process with $PB4$, when the process is not allowed to execute commands (for Webshell only).

$PB5$: If a process has executed shell command, we label the process with $PB5$, for the process may reverse a shell to others.

$PB6$: If a process has modified the file that control scheduled tasks such as $/etc/crontab$, we label the process with $PB6$, for it may cause persistent control.

$PB7$: If a process has modified the file that control permissions such as $/etc/sudoers$, we label the process with $PB7$, for it may cause privilege escalation.

$PB8$: If a process has read high-value information, we label the process with $PB8$, for it may cause data exfiltration.

\subsection{Explanation of Defined Label in Table~\ref{tab:file_label}}
\label{exp:file_label}

$FU1$: If a file is uploaded, we label the file with $FU1$, for it can be untrustworthy.

$FU2$: If a file contains data from the network, we label the file with $FU2$, for the untrusted data may cause an attack later (code execution).

% FS2: If a file was written by a process that executed dangerous code, we label it FS2.

% FS3: If a file is set as sensitive by us, we label it FS3, for all processes which want to access it should be labeled.

$FU3$: If a file does not exist, we label it with $FU3$. In a Living-off-the-land attack, accessing to a non-existent file is marked as access to a $(null)$ file.

$FU4$: If a file is written by the Webshell attack, we label the file with $FU4$, in order to capture the attack chain with webshell as the entry point.

$FU5$: If a file is written by the RAT attack, we label the file with $FU5$, in order to capture the attack chain with RAT as the entry point.

$FU6$: If a file is written by the Living-off-the-land attack, we label the file with $FU6$, in order to capture the attack chain with Living-off-the-land as the entry point.

$FH1$: If a file can control scheduled tasks such as $/etc/crontab$, we label it with $FH1$.

$FH2$: If a file can control permissions such as $/etc/sudoers$, we label it with $FH2$.

$FH3$: If a file holds sensitive information such as $/etc/passwd$, we label it with $FH3$.

$FH4$: If a file saves historical commands such as $.bash\_history$, we label it with $FH4$.

$FH5$: If a file is written by the process that have read sensitive information, we label the file with $FH5$, for it may cause data exfiltration.

\subsection{Event Used in \sn{} and Explanation}
\label{exp:event_used}

\begin{table}[h]
\centering
\caption{Event definition used in \sn{}.}
\label{tab:event_label}
\begin{tabular}{|c|c|}
\hline
\textbf{Events} & \textbf{Description}\\ \hline
E0 & File Read \\ \hline
E1 & File Write \\ \hline
E2 & Fork \\ \hline
E3 & Execute \\ \hline
E4 & LoadLibrary \\ \hline
E5 & File Delete \\ \hline
E6 & File Rename \\ \hline
E7 & File Create \\ \hline
E8 & File Property \\ \hline
E9 & Exit \\ \hline
E10 & LoadElf \\ \hline
E11 & File Open \\ \hline
E12 & File Close \\ \hline 
E13 & Fork with shared open file \\ \hline
E14 & File Open with close-on-exec mark \\ \hline
E15 & File Mmap \\ \hline
\end{tabular}
\end{table}

\subsection{Explanation of Defined Transfer Rules in Table~\ref{tab:Transfer_Rules}}
\label{exp:Transfer_Rules}

$PS1$-E1-$FU2$-D: If a process with label $PS1$ writes a file, we label the file with $FU2$.

$PS3$-E0-$FU2$-R: If a file with label $FU2$ is accessed by a process, we label the process with $PS3$.

$PS3$-E1-$FU2$-D: If a process with label $PS3$ writes a file, we label the file with $FU2$.

$PS4$-E0/E15-$FU1$-R: If a file with label $FU1$ is read or loaded by a process, we label the process with $PS4$.

$PS5$-E0/E10/E15-$FU3$-R: If a file with label $FU3$ is interacted by a process, we label the process with $PS5$.

$PB1$-E3-$FU2$-R: If a file with label $FU2$ is executed by a process, we label the process with $PB1$.

$PB1$-E10-$FU2$-R: If a file with label $FU2$ is loaded by a process, we label the process with $PB1$.

$PB1$-E1-$FU2$-D: If a process with label $PB1$ writes a file, we label the file with $FU2$.

$PB4$-E1-$FU4$-D: If a process with label $PB4$ writes a file, we label the file with $FU4$.

$PS4$-E0/E15-$FU4$-R: If a file with label $FU4$ is read or loaded by process, we label the process with $PS4$.

$PB1$-E1-$FU5$-D: If a process with label $PB1$ writes a file, we label the file with $FU5$.

$PB1$-E0-$FU5$-R: If a file with label $FU5$ is accessed by a process, we label the process with $PB1$.

$PS5$-E1-$FU6$-D: If a process with label $PS5$ writes a file, we label the file with $FU6$.

$PS5$-E0-$FU6$-R: If a file with label $FU6$ is accessed by a process, we label the process with $PS5$.

$PB6$-E1-$FH1$-R: If a file with $FH1$ is written by a process, we label the process with $PB6$.

$PB7$-E1-$FH2$-R: If a file with $FH2$ is written by a process, we label the process with $PB7$.

$PS6$-E0-$FH3$-R: If a file with $FH3$ is accessed by a process, we label the process with $PS6$.

$PS7$-E0-$FH4$-R: If a file with $FH4$ is accessed by a process, we label the process with $PS7$.

$PB6-7|PS6-7$-E1-$FH5$-D: If a process with any one of labels $PS1~4$ writes a file, we label the file with $FH5$.

$PB8$-E0-$FH5$-R: If a file with label $FH5$ is accessed by a process, we label the process with $PB8$.

\subsection{Explanation of Defined Judge Rules in Table~\ref{tab:judge_rules}}
\label{exp:judge_rules}

Download\&Execution: If a process has a network connection and downloads files from the network, executing the file indicates that a threat of Download\&Execution may have occurred. The label $PB1$ represents that a process has the semantics of Download\&Execution.

Webshell: If a process reads/loads the uploaded file and executes any commends without being allowed. It indicates that a Webshell attack may have occurred. The label $PB1$ represents that a process has the semantics of Webshell attack.

RAT: If a process executes a file downloaded from the network and executes a shell command, which reverses a shell to others, it indicates that a RAT attack may have occurred.  The label $PB1$\&$PB5$ represents that a process has the semantics of RAT attack.

Living-off-the-land: If a process interacts with a file that does not exist, and executes a shell command to reverse the shell to others, it indicates that a Living-off-the-land attack may have occurred. The label $PS5$\&$PB5$ represents that a process has the semantics of Living-off-the-land attack.

Suspicious Behavior: If a process reads or writes the file that controls permissions or holds sensitive information, it indicates that suspicious Behaviors (i.e., persistent stronghold, privilege escalation, credential access, information collection, etc.) may have occurred. The label $PB6|PB7|PS6|PS7$ represents that a process has the semantics of suspicious Behaviors.

Data Exfiltraion: If a process causes suspicious behavior and sends data out, it may cause the threat of data leakage. The label $PB8$ represents that a process has the semantics of data exfiltraion.

APT: Satisfy any one of Webshell attacks, RAT attacks, Living-off-the-land attacks, and satisfies data exfiltraion at the same time. In the case of such an attack chain, APT attacks may exist. The label $((PS4\&PB4)|(PB1\&PB5)|(PS5\&PB5))\&PB8$ represents that a process has the semantics of APTs.

\subsection{Details of Attacks from Laboratory}
\label{exp:lab_attacks}

\textbf{L-1: Webshell attack from the backdoored Apache}. In this attack, port 80 is opened on the host to provide web services, and the website has file upload vulnerabilities while any file can be uploaded and accessed. The attacker successfully obtains the Webshell by uploading the Trojan horse file and accessing and connecting through Ant Sword. But the Webshell only has the www-data user permission. In the subsequent penetration, it is found that in the /etc/crontab file, the administrator uses the script $cleanup.sh$ to clean up the folder where uploaded files are stored once an hour. By viewing the script $cleanup.sh$, the attacker finds that anyone has the permission to modify it, so the attacker adds an attack command in $cleanup.sh$ to get the reverse shell. Then the attacker starts to monitor and wait for the script to be executed by the crontab. Soon the attacker obtained the permission of the administrator. By using $id$ command, the attacker finds that the administrator belongs to the root group and has all permissions required by the attacker.

Through the shell of the administrator, the attacker can modify /etc/crontab and /etc/sudoers to keep persistent and exploit benign users. Also, the attacker can check /etc/passwd and .bash\_history to obtain other user's credentials and collect their information. After that, the attacker writes the valuable information into a file and finally sends.

\textbf{L-2: Remote Access Trojan from the phishing website}. In this attack, the attacker uses a phishing website to put the Trojan file into the target host. After the monitoring is turned on and the Trojan file is executed, the permission of benign user are obtained. Afterwards, same as the follow-up operation in the Webshell attack scenario, the timing script in the crontab is found, and the script is modified to obtain higher user's permissions. After that, the sensitive file is modified to achieve the purpose of persistence control and escalation. The sensitive file and user credentials are read, written into the file and finally sends.

\textbf{L-3: Living-off-the-land attack from vulnerable service}. In this attack, the attacker finds that a special service is running on port 29273 on the host, and the permission of users could be obtained through interaction of string overflow. The subsequent attack scenario is similar to L-1 and L-2.

\subsection{Details of Attacks from Darpa Engagement}
\label{exp:darpa_attacks}

% \subsubsection{The Dataset-A from Darpa Engagement}

% The datasets of this experiment contains multiple attacks. The attacks in this data set will be described below.

\textbf{E-1: Information gather and exfiltration}. In this attack, the attacker uses the username and password of a target user to log into the system and collect sensitive information, reading $/etc/shadow$, $/etc/passwd$ files, executing commands such as $ifconfig$, $tcpdump$, $ps aux$, $groups$, $dirname$, $etc.$, and returning the results to display on $/dev/pts/1$.

\textbf{E-1: Malicious file download and execute}. In this attack, the $scp -t ./ccleaner$ command is used to download a file named $ccleaner$ from an unknown address to connect to the external IP $64.95.25.213$, then to modify $SrcSinkObject$. The file $ccleaner$ is executed to obtain permissions and collect sensitive information from the system. The process also forks a $dbus daemon$ process, which reads files such as $/proc/filesystem$, $/proc/mount$, and $/etc/passwd$. After that, the $ccleaner$ and the files generated by it are moved laterally to another host in the intranet through $scp$.

Next a file named $hc$ is copied from the external IP $108.25.125.188$ to the host, and then the file is executed. In this execution, the file loads a large number of binary and library files, and reads a large number of $/proc/directory$ files including $/proc/net/directory$ files and $/etc/passwd$ and other sensitive files. After that, the file is connected to the external IP, and a file named $/tmp/ext96481$ is generated. The file is subsequently executed, the permissions were elevated, and the sensitive file $/etc/passwd$ is read.

% \subsubsection{The Dataset-B from Darpa Engagement}

\textbf{E-2: In-memory attack with firefox}. In this attack, there is a writable and executable memory space in the $firefox$ process. The process also modifies the malicious device file $/dev/glx\_alsa\_675$ and increases its permissions. Next, when the $firefox$ process communicates with the external IP $86.129.31.201$, a malicious library file named $/tmp/libnet.so$ is dropped. Then the $sshd$ process loads the malicious file to connect with the IP address. Subsequently, the malicious $sshd$ process extracts multiple files from the host and transfers these files to the attacker's host. Finally, the process creates an executable file named $/home/admin/files/docs/audiobackup$ and executes it.

% Figure Block
\end{document}